%% file: report.tex
\documentclass[12pt,dvips,a4paper]{article}
\usepackage[pdftex]{graphicx}
\newcommand{\sbf}[1]{\boldsymbol{#1}}
\newcommand{\mbf}[1]{\mathbf{#1}}

\newcommand{\veps}{\varepsilon}

\usepackage[pdftex]{graphicx,xcolor}

\usepackage{amssymb}
\usepackage{amsmath}
\usepackage{setspace}
\usepackage[round]{natbib}

\renewcommand{\title}{CDPM2: A damage-plasticity approach to modelling the failure of concrete}

\begin{document}

\begin{center} \begin{LARGE} \textbf{\title} \end{LARGE} \end{center}

\begin{center} Peter Grassl$^{1*}$, Dimitrios Xenos$^{1}$, Ulrika Nystr\"{o}m$^{2}$, Rasmus Rempling$^{2}$, Kent Gylltoft$^{2}$\end{center}

$^{1}$School of Engineering, University of Glasgow, Glasgow, UK\\
$^{2}$Department of Civil and Environmental Engineering, Chalmers University of Technology, G\"{o}teborg, Sweden\\

$^*$Corresponding author: Email: peter.grassl@glasgow.ac.uk, Phone: +44 141 330 5208, Fax: +44 141 330 4907\\

\section*{Abstract}

A constitutive model based on the combination of damage mechanics and plasticity is developed to analyse the failure of concrete structures.
The aim is to obtain a model, which describes the important characteristics of the failure process of concrete subjected to multiaxial  loading. 
This is achieved by combining an effective stress based plasticity model with a damage model based on plastic and elastic strain measures.
The model response in tension, uni-, bi- and triaxial compression is compared to experimental results.
The model describes well the increase in strength and displacement capacity for increasing confinement levels.
Furthermore, the model is applied to the structural analyses of tensile and compressive failure. 

Keywords: concrete; constitutive model; plasticity; damage mechanics; fracture; mesh dependence

\section{Introduction}

Concrete is a strongly heterogeneous material, which exhibits a complex nonlinear mechanical behaviour.
Failure in tension and low confined compression is characterised by softening which is defined as decreasing stress with increasing deformations.
This softening response is accompanied by a reduction of the unloading stiffness of concrete, and irreversible (permanent) deformations, which are localised in narrow zones often called cracks or shear bands.
On the other hand, the behaviour of concrete subjected to high confined compression is characterised by a ductile hardening response; that is, increasing stress with increasing deformations.
These phenomena should be considered in a constitutive model for analysing the multiaxial behaviour of concrete structures.

There are many constitutive models for the nonlinear response of concrete proposed in the literature. 
Commonly used frameworks are plasticity, damage mechanics and combinations of plasticity and damage mechanics.
Stress-based plasticity models are useful for the modelling of concrete subjected to triaxial stress states, since the yield surface corresponds at a certain stage of hardening to the strength envelope of concrete \citep{Leo35, Willam74, PraWil89, Etse94, Menetrey95, Piv01, GraLunGyl02, PapKap07, CerPap08, FolEts12}. 
Furthermore, the strain split into elastic and plastic parts represents realistically the observed deformations in confined compression, so that unloading and path-dependency can be described well.
However, plasticity models are not able to describe the reduction of the unloading stiffness that is observed in experiments.
Conversely, damage mechanics models are based on the concept of a gradual reduction of the elastic stiffness \citep{Kac80, Mazars84, Ort85, Res87, MazPij89, CarRizWil01a, TaoPhi05, VoyKat09}.
For strain-based isotropic damage mechanics models, the stress evaluation procedure is explicit, which allows for a direct determination of the stress state, without an iterative calculation procedure. Furthermore, the stiffness degradation in tensile and low confined compressive loading observed in experiments can be described.
However, isotropic damage mechanics models are often unable to describe irreversible deformations observed in experiments and are mainly limited to tensile and low confined compression stress states.
On the other hand, combinations of isotropic damage and plasticity are widely used for modelling both tensile and compressive failure and many different models have been proposed in the literature \citep{Ju89,LeeFen98,JasHuePijGha06,GraJir06,NguHou08,NguKor08,VoyTaqKat08,Gra09b,SanHueOli11,ValHof12}.

One popular class of damage-plastic models relies on a combination of stress-based plasticity formulated in the effective (undamaged) stress space combined with a strain based damage model.
The combined damage-plasticity model recently developed by \citet{GraJir06, GraJir06a} belongs to this group.
This model, called here Concrete Damage Plasticity Model 1 (CDPM1), is characterised by a very good agreement with a wide range of experimental results of concrete subjected to multiaxial stress states.
Furthermore, it has been used in structural analysis in combination with techniques to obtain mesh-independent results and has shown to be robust \citep{GraJir06a,ValHof12}.
However, CDPM1 is based on a single damage parameter for both tension and compression. 
This is sufficient for monotonic loading with unloading, but is not suitable for modelling the transition from tensile to compressive failure realistically. 
When the model was proposed, this limitation was already noted and a generalisation to isotropic formulations with several damage parameters was recommended.
In the present work, CDPM1 is revisted to address this issue by proposing separate damage variables for tension and compression. The introduction of two isotropic damage variables for tension and compression was motivated by the work of \citet{Mazars84, Ort85, FicBorPij99}.
Secondly, in CDPM1, a perfect plastic response in the nominal post-peak regime is assumed for the plasticity part and damage is determined by a function of the plastic strain. 
For the nonlocal version of CDPM1 presented in \citet{GraJir06a}, this perfect-plastic response resulted in mesh-dependent plastic strain profiles, although the overall load-displacement response was mesh-independent. 
Already in \citet{GraJir06a}, it was suggested that the plastic strain profile could be made mesh-independent by introducing hardening in the plasticity model for the nominal post-peak regime.
In the present model, the damage functions for tension and compression depend on both plastic and elastic strain components. Furthermore, hardening is introduced in the nominal post-peak regime. With these extensions, the damage laws can be analytically related to chosen stress-inelastic strain relations, which simplifies the calibration procedure.
The extension to hardening is based on recent 1D damage-plasticity model developments in \citet{Gra09b}, which are here for the first time applied to a 3D model.
The present damage-plasticity model for concrete failure is an augmentation of CDPM1. Therefore, the model is called here CDPM2.
The aim of this article is to present in detail the new phenomenological model and to demonstrate that this model is capable of describing the influence of confinement on strength and displacement capacity, the presence of irreversible displacements and the reduction of unloading stiffness, and the transition from tensile to compressive failure realistically. Furthermore, it will be shown, by analysing structural tests, that CDPM2 is able to describe concrete failure mesh independently.

\section{Damage-plasticity constitutive model}

\subsection{General framework} \label{sec:general}

The damage plasticity constitutive model is based on the following stress-strain relationship:
\begin{equation}\label{eq:general}
\boldsymbol{\sigma} = \left(1-\omega_{\rm t}\right) \bar{\boldsymbol{\sigma}}_{\rm t} + \left(1-\omega_{\rm c}\right) \bar{\boldsymbol{\sigma}}_{\rm c}
\end{equation}
where $\bar{\boldsymbol{\sigma}}_{\rm t}$ and $\bar{\boldsymbol{\sigma}}_{\rm c}$ are the positive and negative parts of the effective stress tensor $\bar{\boldsymbol{\sigma}}$, respectively, and $\omega_{\rm t}$ and $\omega_{\rm c}$ are two scalar damage variables, ranging from 0 (undamaged) to 1 (fully damaged).
The effective stress $\bar{\boldsymbol{\sigma}}$ is defined as
\begin{equation}\label{eq:effectiveStress}
\bar{\boldsymbol{\sigma}} = \mathbf{D}_{\rm e} : \left(\boldsymbol{\varepsilon} - \boldsymbol{\varepsilon}_{\rm p}\right) 
\end{equation}
where $\mathbf{D}_{\rm e}$ is the elastic stiffness tensor based on the elastic Young's modulus $E$ and Poisson's ratio $\nu$, $\boldsymbol{\varepsilon}$ is the strain tensor and $\boldsymbol{\varepsilon}_{\rm p}$ is the plastic strain tensor. The positive and negative parts of the effective stress $\bar{\boldsymbol{\sigma}}$ in (\ref{eq:general}) are determined from the principal effective stress $\bar{\boldsymbol{\sigma}}_{\rm p}$ as $\bar{\boldsymbol{\sigma}}_{\rm pt} = \left< \bar{\boldsymbol{\sigma}}_{\rm p} \right>_+$ and $\bar{\boldsymbol{\sigma}}_{\rm pc} = \left< \bar{\boldsymbol{\sigma}}_{\rm p} \right>_-$, where $\left< \right>_+$ and $\left< \right>_-$ are positive and negative part operators, respectively, defined as $\left< x\right>_+ = \max\left(0,x\right)$ and $\left< x\right>_- = \min\left(0,x\right)$.
 For instance, for a combined tensile and compressive stress state with principal effective stress components $\bar{\boldsymbol{\sigma}}_{\rm p} = \left( - \bar{\sigma}, 0.2 \bar{\sigma}, 0.1 \bar{\sigma} \right)^{\rm T}$, the positive and negative principal stresses are $\bar{\boldsymbol{\sigma}}_{\rm pt} = \left( 0, 0.2 \bar{\sigma}, 0.1 \bar{\sigma} \right)^{\rm T}$ and $\bar{\boldsymbol{\sigma}}_{\rm pc} = \left( -\bar{\sigma}, 0, 0 \right)^{\rm T}$, respectively. 

The plasticity model is based on the effective stress, which is independent of damage. 
The model is described by the yield function, the flow rule, the evolution law for the hardening variable and the loading-unloading conditions. 
The form of the yield function is
\begin{equation} \label{eq:yield}
f_{\rm p} \left(\bar{\boldsymbol{\sigma}}, \kappa_{\rm p} \right) = F \left(\bar{\boldsymbol{\sigma}}, q_{\rm h1}, q_{\rm h2}\right)
\end{equation} 
where $q_{\rm h1}\left(\kappa_{\rm p}\right)$ and $q_{\rm h2}\left(\kappa_{\rm p}\right)$ are dimensionless functions controlling the evolution of the size and shape of the yield surface.
The flow rule is
\begin{equation}\label{eq:flowRule}
\dot{\boldsymbol{\varepsilon}}_{\rm p} = \dot{\lambda} \dfrac{\partial g_{\rm p}}{\partial \bar{\boldsymbol{\sigma}}} \left(\bar{\boldsymbol{\sigma}},\kappa_{\rm p}\right)
\end{equation}
where $\dot{\boldsymbol{\varepsilon}}_{\rm p}$ is the rate of the plastic strain, $\dot{\lambda}$ is the rate of the plastic multiplier and $g_{\rm p}$ is the plastic potential.
The rate of the hardening variable $\kappa_{\rm p}$ is related to the rate of the plastic strain by an evolution law.
The loading-unloading conditions are
\begin{equation}\label{eq:UnloadPlast}
f_{\rm p}\leq 0, \hskip 5mm \dot{\lambda} \geq 0, \hskip 5mm \dot{\lambda} f_{\rm p} = 0
\end{equation}
A detailed description of the individual components of the plasticity part of the model are discussed in Section~\ref{sec:plast}.

The damage part of the model is described by the damage loading functions, loading unloading conditions and the evolution laws for damage variables for tension and compression.
For tensile damage, the main equations are
\begin{equation} \label{eq:tensileLoading}
f_{\rm dt} = \tilde{\varepsilon}_{\rm t}(\bar{\boldsymbol{\sigma}}) - \kappa_{\rm dt}
\end{equation}
\begin{equation} \label{eq:tensileUnloading}
f_{\rm dt}\leq 0 \mbox{,} \hspace{0.5cm} \dot{\kappa}_{\rm dt} \geq 0 \mbox{,} \hspace{0.5cm} \dot{\kappa}_{\rm dt} f_{\rm dt} = 0 
\end{equation}
\begin{equation}
\omega_{\rm t} = g_{\rm dt} \left( \kappa_{\rm dt}, \kappa_{\rm dt1}, \kappa_{\rm dt2}\right) 
\end{equation}
For compression, they are 
\begin{equation} \label{eq:compressiveLoading}
f_{\rm dc} = \alpha_{\rm c}\tilde{\varepsilon}_{\rm c}(\bar{\boldsymbol{\sigma}}) - \kappa_{\rm dc}
\end{equation}
\begin{equation} \label{eq:compressiveUnloading}
f_{\rm dc}\leq 0 \mbox{,} \hspace{0.5cm} \dot{\kappa}_{\rm dc} \geq 0 \mbox{,} \hspace{0.5cm} \dot{\kappa}_{\rm dc} f_{\rm dc} = 0 
\end{equation}
\begin{equation}
\omega_{\rm c} = g_{\rm dc} \left( \kappa_{\rm dc}, \kappa_{\rm dc1}, \kappa_{\rm dc2}\right) 
\end{equation}
Here, $f_{\rm dt}$ and $f_{\rm dc}$ are the loading functions, $\tilde{\varepsilon}_{\rm t}(\bar{\boldsymbol{\sigma}})$ and $\tilde{\varepsilon}_{\rm c}(\bar{\boldsymbol{\sigma}})$ are the equivalent strains and $\kappa_{\rm dt}$, $\kappa_{\rm dt1}$, $\kappa_{\rm dt2}$, $\kappa_{\rm dc}$, $\kappa_{\rm dc1}$ and $\kappa_{\rm dc2}$ are damage history variables. Furthermore, $\alpha_{\rm c}$ is a variable that distinguishes between tensile and compressive loading. A detailed description of the variables is given in Section~\ref{sec:dam}.

\subsection{Plasticity part} \label{sec:plast}
The plasticity part of the model is formulated in a three-dimensional framework with a pressure-sensitive yield surface, hardening and non-associated flow. 
The main components are the yield function, the flow rule, the hardening law and the evolution law for the hardening variable. 

\subsubsection{Yield function} \label{sec:yieldSurface}
The yield surface is described in terms of the cylindrical coordinates in the principal effective stress space (Haigh-Westergaard coordinates), which are the volumetric effective stress 
\begin{equation}
\bar{\sigma}_{\rm V} = \dfrac{I_1}{3}
\end{equation}
the norm of the deviatoric effective stress
\begin{equation}
\bar{\rho} = \sqrt{2 J_2}
\end{equation}
and the Lode angle
\begin{equation}\label{lode}
\bar{\theta} = \tfrac{1}{3}\arccos\left(\dfrac{3\sqrt{3}}{2} \dfrac{J_3}{J_2^{3/2}}\right)
\end{equation}
The foregoing definitions use the first invariant 
\begin{equation}
I_1 = \bar{\boldsymbol{\sigma}}:\boldsymbol{\delta} = \bar{\sigma}_{ij}\delta_{ij}
\end{equation}
of the effective stress tensor $\bar{\boldsymbol{\sigma}}$,
and the second and third invariants 
\begin{eqnarray}
J_2 &=& \tfrac{1}{2}\bar{\mathbf{s}}:\bar{\mathbf{s}}= \tfrac{1}{2}\bar{\mathbf{s}}^2:\boldsymbol{\delta} = \tfrac{1}{2}\bar{s}_{ij}\bar{s}_{ij}
\\
J_3 &=& \tfrac{1}{3}\bar{\mathbf{s}}^3:\boldsymbol{\delta} = \tfrac{1}{3}\bar{s}_{ij}\bar{s}_{jk}\bar{s}_{ki}
\end{eqnarray}
of the deviatoric effective stress tensor $\bar{\mathbf{s}}=\bar{\boldsymbol{\sigma}} -\boldsymbol{\delta}I_1/3$.

The yield function
\begin{equation} \label{eq:yieldSurface}    
\begin{split}
& f_{\rm p}(\bar{\sigma}_{\rm V},\bar\rho,\bar\theta;\kappa_{\rm p})=  \left\{\left[1-q_{\rm{h1}}(\kappa_{\rm p})\right]\left( \frac{\bar{\rho}} {\sqrt{6}f_{\rm c}} + \frac{\bar{\sigma}_{\rm V}} {f_{\rm c}} \right)^2 + \sqrt{\frac{3}{2}} \frac {\bar{\rho}}{f_{\rm c}} \right\}^2 \\
& +m_0 q^2_{\rm{h1}}(\kappa_{\rm p})q_{\rm{h2}}(\kappa_{\rm p}) \left[\frac{\bar{\rho} }{\sqrt{6}f_{\rm c}}r(\cos{\bar{\theta}}) + \frac{\bar{\sigma}_{\rm V}}{f_{\rm c}} \right] - q_{\rm{h1}}^2(\kappa_{\rm p}) q_{{\rm h2}}^2(\kappa_{\rm p})
\end{split}
\end{equation}
depends on the effective stress (which enters in the form of cylindrical coordinates) and on the hardening variable $\kappa_{\rm p}$ (which enters through the dimensionless variables $q_{\rm h1}$ and $q_{\rm h2}$). Parameter $f_{\rm c}$ is the uniaxial compressive strength. For $q_{\rm h2} = 1$, the yield function is identical to the one of CDPM1.

\begin{figure}
\begin{center}
\includegraphics[width=10cm]{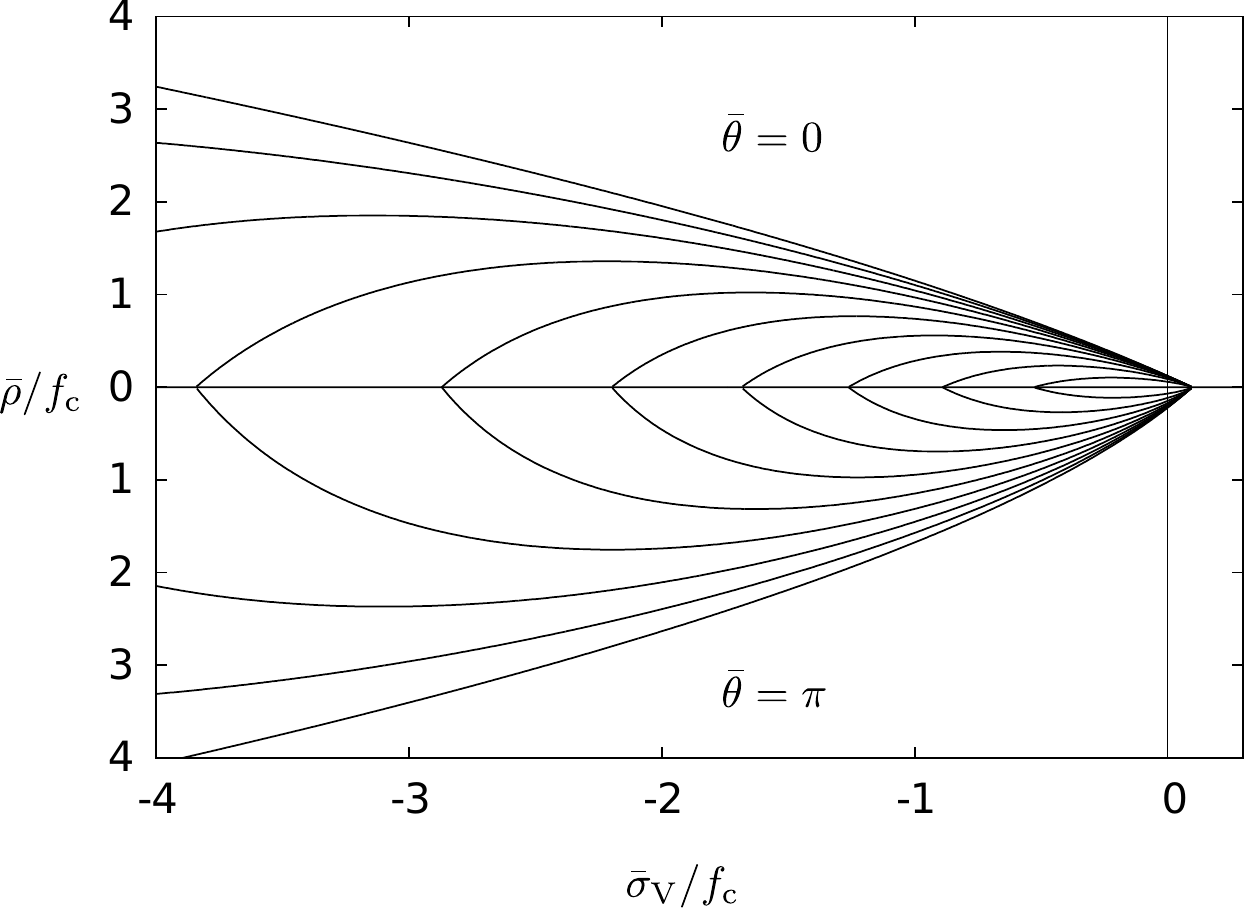}
\end{center}
\caption{The evolution of the meridional section of the yield surface during hardening.}
\label{fig:surfaceMeridian}
\end{figure}

\begin{figure}
\begin{center}
\includegraphics[width=6cm]{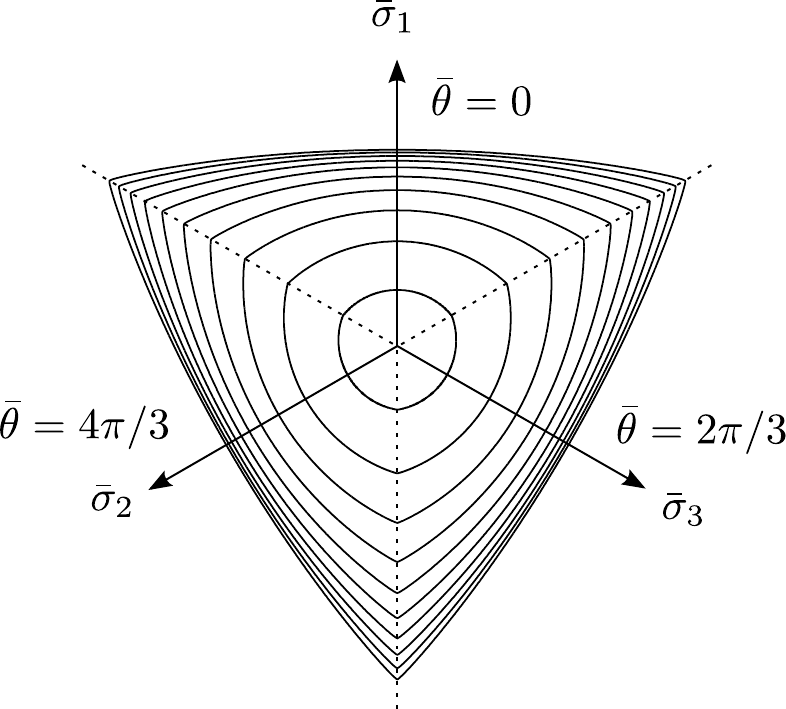}
\end{center}
\caption{The evolution of the deviatoric section of the yield surface during hardening for a constant volumetric stress of $\bar{\sigma}_{\rm V} = - f_{\rm c}/3$.}
\label{fig:surfaceDeviatoric}
\end{figure}

The meridians of the yield surface $f_{\rm p}=0$ are parabolic, and the deviatoric sections change from triangular shapes at low confinement to almost circular shapes at high confinement.
The shape of the deviatoric section is controlled by the function
\begin{equation} \label{eq:rFunction}
\begin{split}
&r(\cos{\bar{\theta}}) = \frac{4(1-e^2)\cos^2{\bar{\theta}} + (2e-1)^2}{2(1-e^2)\cos{\bar{\theta}} + (2e-1)\sqrt{4(1-e^2)\cos^2{\bar{\theta}}+5e^2 -4e}} 
\end{split}
\end{equation} 
proposed by \citet{Willam74}. 
The calibration of the eccentricity parameter $e$ is described in \citet{JirBaz01} and in section~\ref{sec:experiments}.
The friction parameter $m_0$ is given by 
\begin{equation}\label{eq:frictionM}
m_0 = \dfrac{3 \left(f_{\rm c}^2 - f_{\rm t}^2\right)}{f_{\rm c}f_{\rm t}} \dfrac{e}{e+1}
\end{equation}
where $f_{\rm t}$ is the tensile strength.
The shape and evolution of the yield surface is controlled by the variables $q_{\rm h1}$ and $q_{\rm h2}$ (Figs.~\ref{fig:surfaceMeridian}~and~\ref{fig:surfaceDeviatoric}).
If the two variables $q_{\rm h1}$ and $q_{\rm h2}$ in ($\ref{eq:yieldSurface}$) are set equal to one and the resulting yield function is set equal to zero, the failure surface
\begin{equation}\label{eq:strengthEnvelope}
\frac{3}{2} \frac{\bar{\rho}^2}{f_{\rm c}^2} + m_0 \left[\frac{\bar{\rho} }{\sqrt{6}f_{\rm c}}r(\cos{\bar{\theta}}) + \frac{\bar{\sigma}_{\rm V}}{f_{\rm c}} \right] - 1 = 0
\end{equation}
is obtained, which was originally proposed by \citet{Menetrey95}.

\subsubsection{Flow rule}
In the present model, the flow rule in (\ref{eq:flowRule}) is non-associated, which means that the yield function $f_{\rm p}$ and the plastic potential $g_{\rm p}$ do not coincide and, therefore, the direction of the plastic flow $\partial g_{\rm p}/\partial\bar{\boldsymbol{\sigma}}$ is not normal to the yield surface. 
The plastic potential is given as
\begin{equation} \label{eq:plasticPotential}
\begin{split}
& g_{\rm p}(\bar\sigma_{\rm V},\bar\rho;\kappa_{\rm p})=  \left\{\left[1-q_{\rm{h1}}(\kappa_{\rm p})\right] \left( \frac{\bar{\rho}} {\sqrt{6}f_{\rm c}} + \frac{\bar{\sigma}_{\rm V}}{f_{\rm c}} \right)^2 + \sqrt{\frac{3}{2}} \frac {\bar{\rho}}{f_{\rm c}} \right\}^2\\
& + q_{\rm{h1}}^2(\kappa_{\rm p}) \left( \frac{m_0 \bar{\rho}}{\sqrt{6}f_{\rm c}} + \frac{m_{\rm g}(\bar{\sigma}_{\rm V}, \kappa_{\rm p})}{f_{\rm c}} \right)
\end{split}
\end{equation}
where
\begin{equation} \label{eq:mg}
m_{\rm g}(\bar{\sigma}_{\rm V}, \kappa_{\rm p})=A_{\rm g}\left(\kappa_{\rm p}\right)B_{\rm g}\left(\kappa_{\rm p}\right)f_{\rm c} \exp{\frac{\bar{\sigma}_{\rm V} - q_{\rm{h2}}(\kappa_{\rm p}) f_{\rm t}/3}{B_{\rm g}\left(\kappa_{\rm p}\right) f_{\rm c}}}
\end{equation}
is a variable controlling the ratio of volumetric and deviatoric plastic flow.
Here, $A_{\rm g}\left(\kappa_{\rm p}\right)$ and $B_{\rm g}\left(\kappa_{\rm p}\right)$, which depend on $q_{\rm h2}(\kappa_{\rm p})$, are derived from assumptions on the plastic flow in uniaxial tension and compression in the post-peak regime.

The derivation of these two variables is illustrated in the following paragraphs.
Here, the notation $\mathbf{m} \equiv \dfrac{\partial g_{\rm p}}{\partial \bar{\boldsymbol{\sigma}}}$ is introduced.
In the principal stress space, the plastic flow tensor $\mathbf{m}$ has three components, $m_1$, $m_2$ and $m_3$ associated with the three principal stress components.
The flow rule (\ref{eq:flowRule}) is split into a volumetric and a deviatoric part, i.e., the gradient of the plastic potential is decomposed as
\begin{equation}\label{eq:generalFlow}
\mathbf{m} = \dfrac{\partial{g}}{\partial{\bar{\sbf{\sigma}}}} = \frac{\partial{g}}{\partial{\bar{\sigma}_{\rm V}}} \frac{\partial{\bar{\sigma}_{\rm V}}}{\partial{\bar{\sbf{\sigma}}}} + \frac{\partial{g}}{\partial{\bar{\rho}}} \frac{\partial{\bar{\rho}}}{\partial{\bar{\sbf{\sigma}}}}
\end{equation}
Taking into account that $\partial\bar{\sigma}_{\rm V}/\partial\bar{\sbf{\sigma}}=\sbf{\delta}/3$ and $\partial\bar{\rho}/\partial\bar{\sbf{\sigma}}=\bar{\mbf{s}}/\bar\rho$, restricting attention to the post-peak regime (in which $q_{\rm h1} = 1$) and differentiating the plastic potential (\ref{eq:plasticPotential}), we rewrite equation (\ref{eq:generalFlow}) as
\begin{equation}\label{eq:detailedFlow}
\mbf{m} = \dfrac{\partial g}{\partial \bar{\boldsymbol{\sigma}}} =  \frac{{\partial}{m_{\rm g}}}{{\partial}{\bar{\sigma}_{\rm V}}} \frac{\sbf{\delta}}{3f_{\rm c}} 
 + \left(\frac{3}{f_{\rm c}} + \frac{m_0}{\sqrt{6}\bar\rho} \right)\frac{\bar{\mbf{s}}}{f_{\rm c}}
\end{equation}

Experimental results for concrete loaded in uniaxial tension indicate that the strains perpendicular to the loading direction are elastic in the softening regime. Thus, the plastic strain rate in these directions should be equal to zero ($m_2=m_3=0$). 
Under uniaxial tension, the effective stress state in the post-peak regime is characterised by
$\bar{\sigma}_1=f_{\rm t} q_{\rm h2}$, $\bar{\sigma}_2=\bar{\sigma}_3=0$, $\bar{\sigma}_{\rm V}=f_{\rm t} q_{\rm h2}/3$,
$\bar{s}_1=2f_{\rm t} q_{\rm h2}/3$, $\bar{s}_2=\bar{s}_3=-f_{\rm t} q_{\rm h2}/3$ and $\bar\rho=\sqrt{2/3}f_{\rm t} q_{\rm h2}$.
Substituting this into (\ref{eq:detailedFlow}) and enforcing the condition $m_2=m_3=0$, we obtain an equation from which 
\begin{equation}\label{eq:value1}
\frac{{\partial}{m_{\rm g}}}{{\partial}{\bar{\sigma}_{\rm V}}}\left|_{\bar{\sigma}_{\rm V}=f_{\rm t} q_{\rm h2}/3}\right. = \frac{3 f_{\rm t} q_{\rm h2}}{f_{\rm c}} + \frac{m_0}{2}
\end{equation}

In uniaxial compressive experiments, a volumetric expansion is observed in the softening regime. Thus, the inelastic lateral strains are positive while the inelastic axial strain is negative.  In the present approach, a constant ratio $D_{\rm f}=-m_2/m_1=-m_3/m_1$ between lateral and axial plastic strain rates in the softening regime is assumed.
The effective stress state at the end of hardening under uniaxial compression is characterised by
$\bar\sigma_1=-f_{\rm c} q_{\rm h2}$, $\bar\sigma_2=\bar\sigma_3=0$, $\bar\sigma_{\rm V}=-f_{\rm c} q_{\rm h2}/3$,
$\bar{s}_1=-2f_{\rm c} q_{\rm h2}/3$, $\bar{s}_2=\bar{s}_3=f_{\rm c} q_{\rm h2}/3$ and $\bar\rho=\sqrt{2/3}f_{\rm c} q_{\rm h2}$.
Substituting this into  (\ref{eq:detailedFlow}) and enforcing the condition $m_2=m_3=-D_{\rm f}m_1$, we get an equation from which 
\begin{equation}\label{eq:value2}
\frac{{\partial}{m_{\rm g}}}{{\partial}{\bar{\sigma}_{\rm V}}}\left|_{\bar{\sigma}_{\rm V}=-f_{\rm c} q_{\rm h2}/3}\right. = \frac{2D_{\rm f}-1}{D_{\rm f}+1}\left(3 q_{\rm h2}+\frac{m_0}{2}\right)
\end{equation}
Substituting the specific expression for ${\partial}{m_{\rm g}}/{\partial}{\bar{\sigma}_{\rm V}}$ constructed by differentiation of (\ref{eq:mg}) into (\ref{eq:value1}) and (\ref{eq:value2}), we obtain two equations from which parameters  
\begin{eqnarray}\label{eq:ag}
A_{\rm g} &=& \frac{3f_{\rm t} q_{\rm h2}}{f_{\rm c}} + \frac{m_0}{2}
\\
B_{\rm g} &=& \frac{\left(q_{\rm h2}/3\right)\left(1+f_{\rm t}/f_{\rm c}\right)}{\ln A_{\rm g}- \ln\left(2D_{\rm f}-1 \right)- \ln\left(3q_{\rm h2}+m_0/2\right)+ \ln\left(D_{\rm f}+1 \right)}
\label{eq:bg}
\end{eqnarray}
can be computed.
The gradient of the dilation variable $m_{\rm g}$ in (\ref{eq:mg}) decreases with increasing confinement.
The limit $\bar{\sigma}_{\rm V} \rightarrow - \infty$ corresponds to purely deviatoric flow.  
As in CDPM1, the plastic potential does not depend on the third Haigh-Westergaard coordinate (Lode angle $\bar{\theta}$), which increases the efficiency of the implementation and the robustness of the model.

\subsubsection{Hardening law}\label{sec:hardeningLaw}
The dimensionless variables $q_{\rm h1}$ and $q_{\rm h2}$ that appear in (\ref{eq:yieldSurface}), (\ref{eq:plasticPotential}) and (\ref{eq:mg}) are functions of the hardening variable $\kappa_{\rm p}$. They control the evolution of the size and shape of the yield surface and plastic potential. 
The first hardening law $q_{\rm h1}$ is
\begin{equation} \label{eq:hardeningLawOne}
q_{\rm h1}(\kappa_{\rm p}) = 
\left \{ \begin{array}{ll} q_{\rm h0} + \left(1-q_{\rm h0}\right) \left( \kappa_{\rm p}^3 - 3 \kappa_{\rm p}^2 + 3 \kappa_{\rm p} \right) - H_{\rm p} \left(\kappa_{\rm p}^3 - 3 \kappa_{\rm p}^2 + 2 \kappa_{\rm p}\right) & \mbox{if $\kappa_{\rm p} < 1$} \\
1 & \mbox{if $\kappa_{\rm p} \ge 1$}
\end{array}
\right.
\end{equation}
The second hardening law $q_{\rm h2}$ is given by
\begin{equation} \label{eq:hardeningLawTwo}
q_{\rm h2}(\kappa_{\rm p}) = \left \{ \begin{array}{ll} 1 & \mbox{if $\kappa_{\rm p} < 1$} \\
1 + H_{\rm p} (\kappa_{\rm p} - 1)  & \mbox{if $\kappa_{\rm p} \ge 1$}
\end{array}
\right.
\end{equation}

\begin{figure}
\begin{center}
\includegraphics[width=10cm]{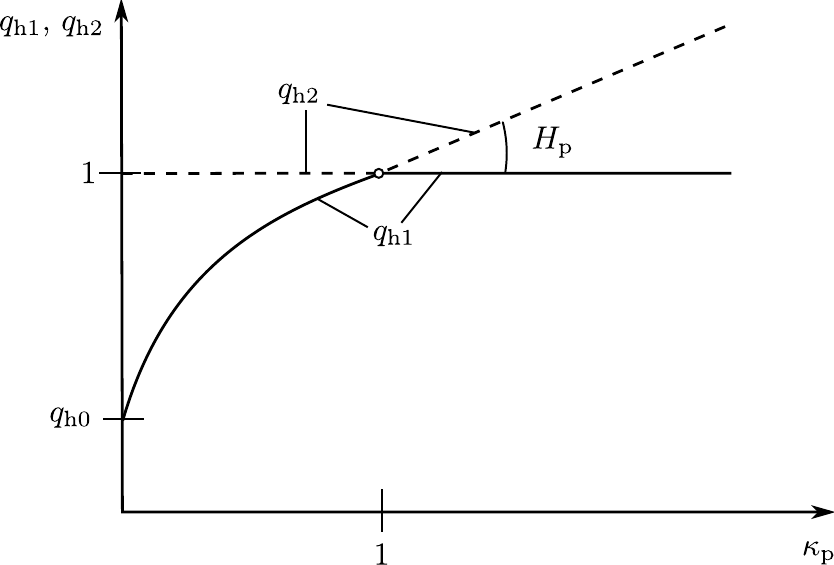}
\end{center}
\caption{The two hardening laws $q_{\rm h1}$ (solid line) and $q_{\rm h2}$ (dashed line).}
\label{fig:hardening}
\end{figure}

The initial inclination of the hardening curve $q_{\rm h1}$ at $\kappa_{\rm p}=0$ is positive and finite, and the inclinations of $q_{\rm h1}$ on the left of $\kappa_{\rm p} = 1$ and $q_{\rm h2}$ on the right of  $\kappa_{\rm p} = 1$ are equal to $H_{\rm p}$, as depicted in Fig.~\ref{fig:hardening}.
For $H_{\rm p} = 0$, the hardening law reduces to the one proposed in \cite{GraJir06}.

\subsubsection{Hardening variable}
The evolution law for the hardening variable,
\begin{equation}\label{eq:dotkappap}
\dot{\kappa}_{\rm {p}} = \frac {\| \dot{\veps_{\rm{p}}}\|}{x_{\rm{h}}\left(\bar{\sigma}_{\rm V} \right)} \left(2 \cos{\bar{\theta}}\right)^2 = \frac {\dot{\lambda} \| \mbf{m}\|}{x_{\rm{h}}\left(\bar{\sigma}_{\rm V} \right)} \left(2\cos{\bar{\theta}}\right)^2
\end{equation}
sets the rate of the hardening variable equal to the norm of the plastic strain rate scaled by a hardening ductility measure 
\begin{equation} \label{eq:hardeningMeasure}
x_{\rm{h}}\left(\bar{\sigma}_{\rm V}\right) = \left\{ \begin{array}{ll}
A_{\rm{h}} - \left(A_{\rm{h}} - B_{\rm{h}} \right)\exp{\left(-R_{\rm h}(\bar{\sigma}_{\rm V})/C_{\rm h}\right)} & \mbox{if $R_{\rm h}(\bar{\sigma}_{\rm V}) \geq 0 $} \\[5mm]
E_{\rm h}\exp({R_{\rm h}(\bar{\sigma}_{\rm V})/F_{\rm h}})+D_{\rm h} & \mbox{if $R_{\rm h}(\bar{\sigma}_{\rm V}) < 0$}
\end{array}
\right.
\end{equation}
For pure volumetric stress states, $\bar{\theta}$ in (\ref{eq:dotkappap}) is set to zero.
The dependence of the scaling factor $x_{\rm{h}}$ on the volumetric stress $\bar{\sigma}_{\rm V}$ is constructed such that the model response is more ductile under compression.
The variable 
\begin{equation}
R_{\rm h}(\bar{\sigma}_{\rm V}) = -\frac{\bar{\sigma}_{\rm V}}{f_{\rm c}}-\frac{1}{3}
\end{equation}
is a linear function of the volumetric effective stress.
Model parameters $A_{\rm h}, B_{\rm h}, C_{\rm h}$ and $D_{\rm h}$ are calibrated from the values of strain at peak stress under uniaxial tension, uniaxial compression and triaxial compression, whereas the parameters $E_{\rm h}$ and $F_{\rm h}$ are determined from the conditions of a smooth transition between the two parts of equation \eqref{eq:hardeningMeasure} at $R_{\rm h} = 0$:
\begin{eqnarray}
E_{\rm h} &=& B_{\rm h} - D_{\rm h}
\\
F_{\rm h} &=& \frac{\left(B_{\rm h} -D_{\rm h}\right)C_{\rm h}}{A_{\rm h}-B_{\rm h}}
\end{eqnarray}
This definition of the hardening variable is identical to the one in CDPM1 described in \citet{GraJir06}, where the calibration procedure for this part of the model is described.
 
\subsection{Damage part} \label{sec:dam}

Damage is initiated when the maximum equivalent strain in the history of the material reaches the threshold $\varepsilon_{0} = f_{\rm t}/E$. For uniaxial tension only, the equivalent strain could be chosen as  $\tilde{\varepsilon} = \bar{\sigma}_{\rm t}/E$, where $\bar{\sigma}_{\rm t}$ is the effective uniaxial tensile stress. 
Thus, damage initiation would be linked to the axial elastic strain.
However, for general triaxial stress states a more advanced equivalent strain expression is required, which predicts damage initiation when the strength envelope is reached.
This expression is determined from the yield surface ($f_{\rm p} = 0$) by setting $q_{\rm {h1}}=1$ and $q_{\rm{h2}} = \tilde{\varepsilon}/\varepsilon_{0}$.
From this quadratic equation for $\tilde{\varepsilon}$, the equivalent strain is determined as
\begin{equation}\label{eq:equivStrain}
\tilde{\varepsilon} = \dfrac{\varepsilon_0 m_0}{2}  \left(\dfrac{\bar{\rho}}{\sqrt{6} f_{\rm c}} r\left(\cos\theta\right) + \dfrac{\bar{\sigma}_{\rm V}}{f_{c}}\right) + \sqrt{\dfrac{\varepsilon_0^2 m_0^2}{4} \left(\dfrac{\bar{\rho}}{\sqrt{6}f_{\rm c}} r\left(\cos\theta\right) + \dfrac{\bar{\sigma}_{\rm V}}{f_{\rm c}}\right)^2 + \dfrac{3 \varepsilon_0^2 \bar{\rho}^2}{2 f_{\rm c}^2}}
\end{equation}
For uniaxial tension, the effective stress state is defined as $\bar{\sigma}_1=\bar{\sigma}_{\rm t}$, $\bar{\sigma}_2=\bar{\sigma}_3=0$, $\bar{\sigma}_{\rm V} = \bar{\sigma}_{\rm t}/3$, $\bar{s}_1 = 2 \bar{\sigma}_{\rm t}/3$, $\bar{s}_2 = \bar{s}_3 = - \bar{\sigma}_{\rm t}/3$, $\bar{\rho} = \sqrt{2/3} \bar{\sigma}_{\rm t}$ and $r(\cos \theta) = 1/e$.
Setting this into (\ref{eq:equivStrain}) and using the definition of $m_0$ in (\ref{eq:frictionM}) gives
\begin{equation}
\tilde{\varepsilon} = \varepsilon_0 \dfrac{\bar{\sigma}_{\rm t}}{f_{\rm t}} = \bar{\sigma}_{\rm t}/E
\end{equation}
which is suitable equivalent strain for modelling tensile failure.
For uniaxial compression, the effective stress state is defined as $\bar{\sigma}_1 = -\bar{\sigma}_{\rm c}$, $\bar{\sigma}_2 = \bar{\sigma}_3 = 0$, $\bar{\sigma}_{\rm V} = -\bar{\sigma}_{\rm c}/3$, $\bar{s}_{1} = -2/3 \bar{\sigma}_{\rm c}$, $\bar{s}_{2} = \bar{s}_{3} = 1/3 \bar{\sigma}_{\rm c}$, $\bar{\rho} = \sqrt{2/3} \bar{\sigma}_{\rm c}$, and $r(\cos \theta) = 1$.
Here, $\bar{\sigma}_{\rm c}$ is the magnitude of the effective compressive stress.
Setting this into (\ref{eq:equivStrain}), the equivalent strain is
\begin{equation}
\tilde{\varepsilon} = \dfrac{\bar{\sigma}_{\rm c} \varepsilon_0}{f_{\rm c}} =  \dfrac{\bar{\sigma}_{\rm c} f_{\rm t}}{E f_{\rm c}}
\end{equation} 
If $\bar{\sigma}_{\rm c} = (f_{\rm c}/f_{\rm t}) \bar{\sigma}_{\rm t}$, the equivalent strain is again equal to the axial elastic strain component in uniaxial tension.
Consequently, the equivalent strain definition in (\ref{eq:equivStrain}) is suitable for both tension and compression, which is very convenient for relating the damage variables in tension and compression to stress-inelastic strain curves.
 
\begin{figure}
\begin{center}
\includegraphics[width=10cm]{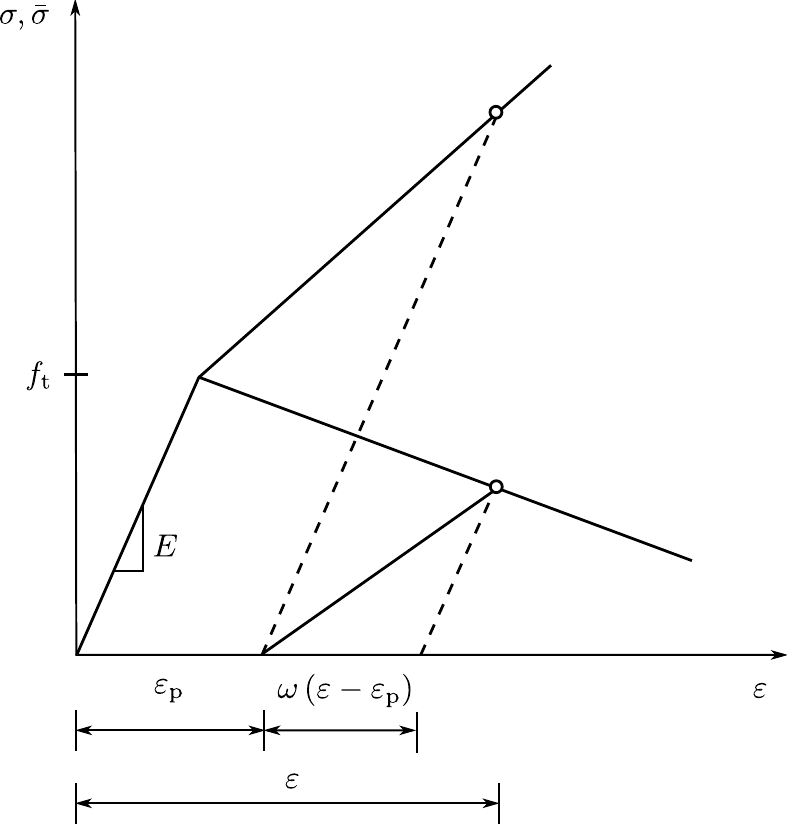}
\end{center}
\caption{Geometrical meaning of the inelastic strain \protect $\varepsilon_{\rm i}$ for the combined damage-plasticity model. The inelastic strain is composed of reversible \protect $\omega \left(\varepsilon - \varepsilon_{\rm p}\right)$ and irreversible $\varepsilon_{\rm p}$ parts. The dashed lines represent elastic unloading with the same stiffness as the initial elastic loading.}
\label{fig:compExplain}
\end{figure}

The damage variables $\omega_{\rm t}$ and $\omega_{\rm c}$ in (\ref{eq:general}) are determined so that a prescribed stress-inelastic strain relation in uniaxial tension is obtained. Since, the damage variables are evaluated for general triaxial stress states, the inelastic strain in uniaxial tension has to be expressed by suitable scalar history variables, which are obtained from total and plastic strain components.
To illustrate the choice of these components, a 1D damage-plastic stress-strain law of the form 
\begin{equation}\label{eq:general1D}
\sigma = \left(1-\omega\right) \bar{\sigma} = \left(1-\omega\right) E  \left( \varepsilon - \varepsilon_{\rm p} \right)
\end{equation}
is considered. Here, $\omega$ is the damage variable.
This law can also be written as
\begin{equation} \label{eq:inelastic}
\sigma = E  \left\{\varepsilon - \left[\varepsilon_{\rm p} + \omega \left(\varepsilon-\varepsilon_{\rm p}\right)\right]\right\} = E \left(\varepsilon - \varepsilon_{\rm i}\right)
\end{equation}
where $\varepsilon_{\rm i}$ is the inelastic strain which is subtracted from the total strain.
The geometrical interpretation of the inelastic strain and its split for monotonic uniaxial tension, linear hardening plasticity and linear damage evolution are shown in Fig.~\ref{fig:compExplain}. Furthermore, the way how the hardening influences damage and plasticity dissipation has been discussed in \citet{Gra09b}.
The part $\omega \left(\varepsilon - \varepsilon_{\rm p}\right)$ is reversible and $\varepsilon_{\rm p}$ is irreversible.
The damage variable is chosen, so that a softening law is obtained, which relates the stress to the inelastic strain, which is written here in generic form as
\begin{equation}\label{eq:genericSoftening}
\sigma = f_{\rm s} \left(\varepsilon_{\rm i}\right)
\end{equation} 
Setting (\ref{eq:inelastic}) equal with (\ref{eq:genericSoftening}) allows for determining the damage variable $\omega$.

However, the inelastic strain $\varepsilon_{\rm i}$ in (\ref{eq:inelastic}) and (\ref{eq:genericSoftening}) needs to be expressed by history variables, so that the expression for the damage variable can be used for non-monotonic loading. Furthermore, to be able to describe also the influence of multiaxial stress states on the damage evolution, the inelastic strain in (\ref{eq:genericSoftening}) is replaced by different history variables than the inelastic strain in (\ref{eq:inelastic}). The choice of the history variables for tension and compression is explained in sections.~\ref{sec:historyTension}~and~\ref{sec:historyCompression}.

\subsubsection{History variables for tension} \label{sec:historyTension}
The tensile damage variable $\omega_{\rm t}$ in (\ref{eq:general}) is defined by three history variables $\kappa_{\rm dt}$, $\kappa_{\rm dt1}$ and $\kappa_{\rm dt2}$. The variable $\kappa_{\rm dt}$ is used in the definition of the inelastic strain in (\ref{eq:inelastic}), while $\kappa_{\rm dt1}$ and $\kappa_{\rm dt2}$ enter the definition of the inelastic strain in (\ref{eq:genericSoftening}).
The history variable $\kappa_{\rm dt}$ is determined from $\tilde{\varepsilon}_{\rm t}$ using (\ref{eq:tensileLoading}) and (\ref{eq:tensileUnloading}).
Here, $\tilde{\varepsilon}_{\rm t}$ is given implicitly in incremental form by
\begin{equation}
\dot{\tilde{\varepsilon}}_{\rm t} = \dot{\tilde{\varepsilon}}
\end{equation}
with $\tilde{\varepsilon}$ given in (\ref{eq:equivStrain}).
For $\kappa_{\rm dt1}$, the inelastic strain component related the plastic strain $\varepsilon_{\rm p}$ is replaced by
\begin{equation}\label{eq:kappaDT1}
\dot{\kappa}_{\rm dt1} = \left \{ \begin{array}{ll}
\dfrac{1}{x_{\rm s}} \|\dot{\boldsymbol{\varepsilon}}_{\rm p}\| & \mbox{if $\dot{\kappa}_{\rm dt} > 0$ and $\kappa_{\rm dt} >\varepsilon_0$}\\
0 & \mbox{if $\dot{\kappa}_{\rm dt} = 0$ or $\kappa_{\rm dt} <\varepsilon_0$}
\end{array} \right.
\end{equation}
Here, the pre-peak plastic strains do not contribute to this history variable, since $\dot{\kappa}_{\rm dt1}$ is only nonzero, if $\kappa_{\rm dt}>\varepsilon_0$.
Finally, the third history variable is related to $\kappa_{\rm dt}$ as
\begin{equation}\label{eq:kappaDT2}
\dot{\kappa}_{\rm dt2} = \dfrac{\dot{\kappa}_{\rm dt}}{x_{\rm s}}
\end{equation}

In (\ref{eq:kappaDT1}) and (\ref{eq:kappaDT2}), $x_{\rm s}$ is a ductility measure,  which describes the influence of multiaxial stress states on the softening response, see Sec.~\ref{sec:damageDuctility}.

\subsubsection{History variables for compression} \label{sec:historyCompression}
The compression damage variable $\omega_{\rm c}$ is also defined by three history variables $\kappa_{\rm dc}$, $\kappa_{\rm dc1}$ and $\kappa_{\rm dc2}$. Analogous to the tensile case, the variable $\kappa_{\rm dc}$ is used in the definition of the inelastic strain in (\ref{eq:inelastic}), while $\kappa_{\rm dc1}$ and $\kappa_{\rm dc2}$ enter the definition of the equivalent strain in (\ref{eq:genericSoftening}).
In addition, a variable  $\alpha_{\rm c}$ is introduced which distinguishes tensile and compressive stresses.
It has the form
\begin{equation} \label{eq:alpha}
\alpha_{\rm c} = \sum_{i=1}^3 \dfrac{\bar{\sigma}_{\rm{pc} i}\left( \bar{\sigma}_{\rm{pt} i} + \bar{\sigma}_{\rm{pc} i}\right)}{\| \bar{\boldsymbol{\sigma}}_{\rm{p}} \|^2}
\end{equation}
where $\bar{\sigma}_{\rm{pt}i}$ and $\bar{\sigma}_{\rm{pc}i}$ are the components of the compressive and tensile part of the principal effective stresses, respectively, which were previously used for the general stress strain law in (\ref{eq:general}).
The variable $\alpha_{\rm c}$ varies from 0 for pure tension to 1 for pure compression. For instance, for the mixed tensile compressive effective stress state $\bar{\boldsymbol{\sigma}}_{\rm p} = \left\{-\bar{\sigma}, 0.2 \bar{\sigma}, 0.1 \bar{\sigma}\right\}$, considered in Sec.~\ref{sec:general}, the variable is $\alpha_{c} = 0.95$.

The history variable $\kappa_{\rm dc}$ is determined from $\tilde{\varepsilon}_{\rm c}$ using (\ref{eq:compressiveLoading}) and (\ref{eq:compressiveUnloading}), where, analogous to the tensile case, the $\varepsilon_{\rm c}$ is specified implicitly by
\begin{equation}
\dot{\tilde{\varepsilon}}_{\rm c} = \alpha_{\rm c} \dot{\tilde{\varepsilon}}
\end{equation}
The other two history variables are
\begin{equation}\label{eq:kappaDC1}
\dot{\kappa}_{\rm dc1} = \left \{ \begin{array}{ll}
\dfrac{\alpha_{\rm c} \beta_{\rm c}}{x_{\rm s}} \|\dot{\boldsymbol{\varepsilon}}_{\rm p}\| & \mbox{if $\dot{\kappa}_{\rm dt} > 0$ $\land$ $\kappa_{\rm dt} >\varepsilon_0$}\\
0 & \mbox{if $\dot{\kappa}_{\rm dt} = 0$ $\lor$ $\kappa_{\rm dt} <\varepsilon_0$}
\end{array} \right.
\end{equation}
and
\begin{equation}\label{eq:kappaDC2}
\dot{\kappa}_{\rm dc2} = \dfrac{\dot{\kappa}_{\rm dc}}{x_{\rm s}}
\end{equation}
In (\ref{eq:kappaDC1}), the factor $\beta_{\rm c}$ is
\begin{equation}
\beta_{\rm c} = \dfrac{f_{\rm t} q_{\rm h2} \sqrt{2/3}}{\bar{\rho} \sqrt{1+2 D_{\rm f}^2}}
\end{equation}
This factor provides a smooth transition from pure damage to damage-plasticity softening processes, which can occur during cyclic loading, as described in section~\ref{sec:cyclic}. 

\subsubsection{Damage variables for bilinear softening} \label{sec:bilinear}
\begin{figure}
\begin{center}
\includegraphics[width=6cm]{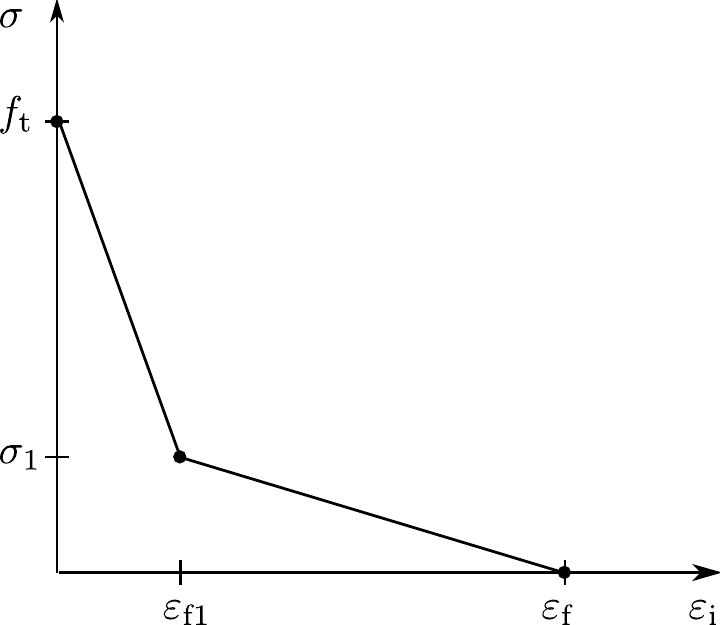}
\end{center}
\caption{Bilinear softening.}
\label{fig:damBilinear}
\end{figure}

With the history variables defined in the previous two sections, the damage variables for tension and compression are determined.
The form of these damage variables depends on the type of softening law considered.
For bilinear softening used in the present study, the stress versus inelastic strain in the softening regime is 
\begin{equation}
\sigma = \left \{ \begin{array}{ll}\label{eq:bilinearCrackOpening}
f_{\rm t} - \dfrac{f_{\rm t} - \sigma_{1}}{\varepsilon_{\rm f1}}\varepsilon_{\rm i} & \mbox{if $0 < \varepsilon_{\rm i} \leq \varepsilon_{\rm f1}$} \vspace{0.2cm} \\
\sigma_{\rm 1} - \dfrac{\sigma_{1}}{\varepsilon_{\rm f}-\varepsilon_{\rm f1}}\left(\varepsilon_{\rm i} - \varepsilon_{\rm f1}\right) & \mbox{if $\varepsilon_{\rm f1} < \varepsilon_{\rm i} \leq \varepsilon_{\rm f}$} \vspace{0.2cm}\\
0 & \mbox{if $\varepsilon_{\rm f} \leq \varepsilon_{\rm i}$}
\end{array}
\right.
\end{equation}
where $\varepsilon_{\rm f}$ is the inelastic strain threshold at which the uniaxial stress is equal to zero and $\varepsilon_{\rm f1}$ is the threshold where the uniaxial stress is equal to $\sigma_{1}$ as shown in Fig.~\ref{fig:damBilinear}.
Furthermore, $\varepsilon_{\rm i}$ is the inelastic strain in the post-peak regime only.
Since damage is irreversible, the inelastic strain $\varepsilon_{\rm i}$ in (\ref{eq:bilinearCrackOpening}) is expressed by irreversible damage history variables as
\begin{equation}\label{eq:inelasticHistory}
\varepsilon_{\rm i} = \kappa_{\rm dt1}+ \omega_{\rm t} \kappa_{\rm dt2}
\end{equation}
Furthermore, the term $\varepsilon-\varepsilon_{\rm p}$ in (\ref{eq:general1D})  is replaced by $\kappa_{\rm dt}$, which gives
\begin{equation}\label{eq:linearStressStrain}
\sigma = \left(1-\omega_{\rm t}\right) E \kappa_{\rm dt}
\end{equation}
Setting (\ref{eq:bilinearCrackOpening}) with (\ref{eq:inelasticHistory}) equal to (\ref{eq:linearStressStrain}), and solving for $\omega_{\rm t}$ gives
\begin{equation}\label{eq:damTenBilinear}
\omega_{\rm t} = \left \{ \begin{array}{ll}
\dfrac{(E \kappa_{\rm dt} -f_{\rm t}) \varepsilon_{\rm f1} - (\sigma_1 - f_{\rm t}) \kappa_{\rm dt1}}{E \kappa_{\rm dt} \varepsilon_{\rm f1} + (\sigma_{1} - f_{\rm t}) \kappa_{\rm dt2}} & \mbox{if $0 < \varepsilon_{\rm i} \leq \varepsilon_{\rm f1}$}\vspace{0.2cm}\\
\dfrac{E \kappa_{\rm dt} \left(\varepsilon_{\rm f} - \varepsilon_{\rm f1}\right) + \sigma_{\rm 1} \left(\kappa_{\rm dt1} - \varepsilon_{\rm f}\right)}{E \kappa_{\rm dt} \left(\varepsilon_{\rm f} - \varepsilon_{\rm f1}\right) - \sigma_{1} \kappa_{\rm dt2}} & \mbox{if $\varepsilon_{\rm f1} < \varepsilon_{\rm i} \leq \varepsilon_{\rm f}$}\vspace{0.2cm}\\
0  & \mbox{if $\varepsilon_{\rm f} < \varepsilon_{\rm i}$}\\
\end{array}
\right.
\end{equation}

For the compressive damage variable, an evolution based on an exponential stress-inelastic strain law is used. The stress versus inelastic strain in the softening regime in compression is
\begin{equation}
\begin{array}{ll}
\sigma = f_{\rm t} \exp\left(-\dfrac{\varepsilon_{\rm i}}{\varepsilon_{\rm fc}}\right) & \mbox{if $0 < \varepsilon_{\rm i}$}
\end{array}
\end{equation}
where $\varepsilon_{\rm fc}$ is an inelastic strain threshold which controls the initial inclination of the softening curve. The use of different damage evolution for tension and compression is one important improvement over CDPM1 as it will shown later on when the structural applications are discussed.

\subsubsection{Ductility measure} \label{sec:damageDuctility}
The history variables $\kappa_{\rm dt1}$, $\kappa_{\rm dt2}$, $\kappa_{\rm dc1}$ and $\kappa_{\rm dc2}$ in (\ref{eq:kappaDT1}), (\ref{eq:kappaDT2}), (\ref{eq:kappaDC1}) and (\ref{eq:kappaDC2}), respectively, depend on a ductility measure $x_{\rm s}$, which takes into account the influence of multiaxial stress states on the damage evolution.
This ductility measure is given by
\begin{equation}
x_{\rm s} = 1+ \left(A_{\rm s}-1\right) R_{\rm s}
\end{equation}
where $R_{\rm s}$ is
\begin{equation}
R_{\rm s} = \left \{ \begin{array}{ll} -\dfrac{\sqrt{6} \bar{\sigma}_{\rm V}}{\bar{\rho}} & \mbox{if $\bar{\sigma}_{\rm V} \leq 0$} \\
    0 & \mbox{if $\bar{\sigma}_{\rm V} > 0$}  \end{array} \right.
\end{equation}
and $A_{\rm s}$ is a model parameter.
For uniaxial compression $\bar{\sigma}_{\rm V}/\bar{\rho} = -1/\sqrt{6}$, so that $R_{\rm s} = 1$ and $x_{\rm s} = A_{\rm s}$, which simplifies the calibration of the softening response in this case.

\subsubsection{Constitutive response to cyclic loading} \label{sec:cyclic}

\begin{figure}
\begin{center}
\includegraphics[width=10cm]{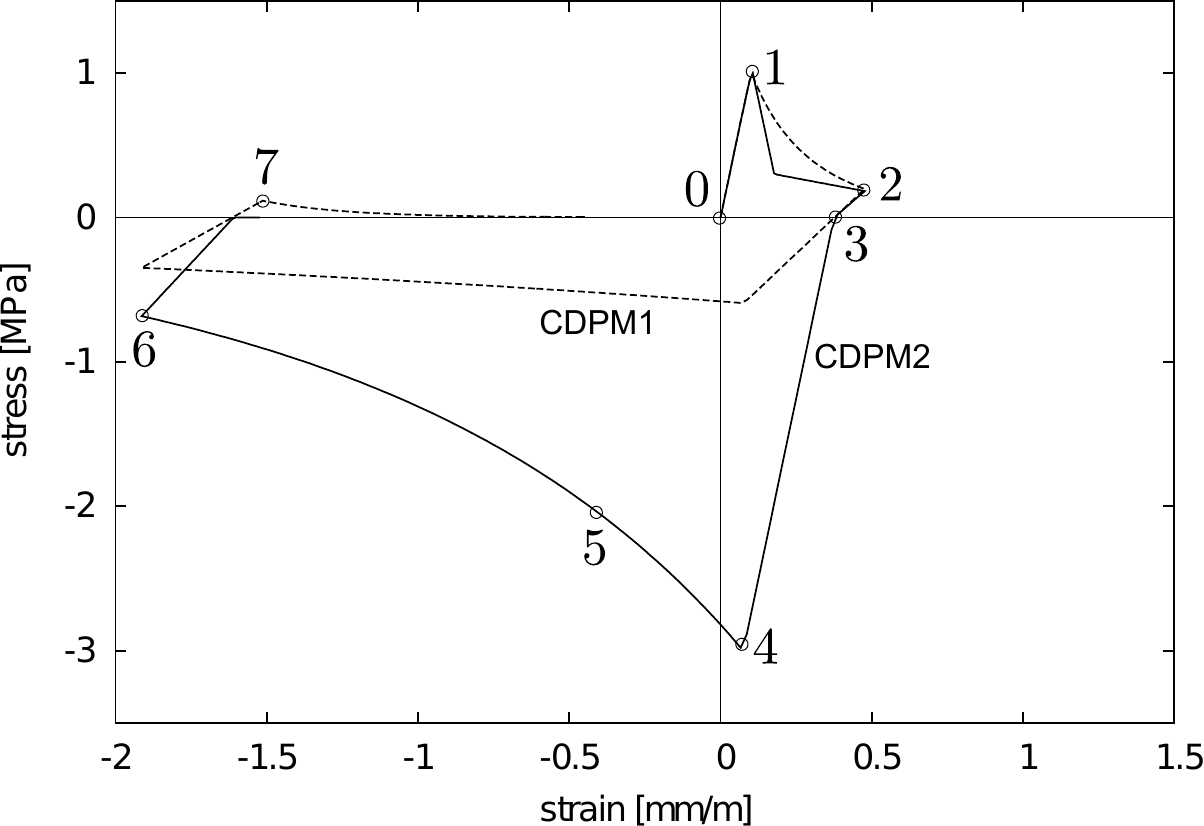}
\end{center}
\caption{Model response for cyclic loading with $f_{\rm t} = 1$ and $f_{\rm c} = 3$ for CDPM2 (solid line) and CDPM1 (dashed line).}
\label{fig:Cyclic}
\end{figure}

The response of the constitutive model is illustrated by a quasi-static strain cycle (Fig.~\ref{fig:Cyclic}, solid line), before it is compared to a wide range of experimental results in the next section.
The strain is increased from point 0 to point 1, where the tensile strength of the material is reached.
Up to point 1, the material response is elastic-plastic with small plastic strains.
With a further increase of the strain from point 1 to point 2, the effective stress part continues to increase, since $H_{\rm p}>0$, whereas the nominal stress decreases, since the tensile damage variable $\omega_{\rm t}$ increases.
A reverse of the strain at point 2 results in an reduction of the stress with an unloading stiffness, which is less than the elastic stiffness of an elasto-plastic model, but greater than the stiffness of an elasto-damage mechanics model, i.e. greater than the secant stiffness.
At point 3, when the stress is equal to zero, a further reduction of the strain leads to a compressive response following a linear stress-strain relationship between the points 3 and 4 with the original Young's modulus $E$ of the undamaged material.
This change of stiffness is obtained by using two damage variables, $\omega_{\rm t}$ and $\omega_{\rm c}$.
At point 3, $\omega_{\rm t}>0$, but $\omega_{\rm c} = 0$.
Up to point 5, no further plastic strains are generated, since the hardening from point 0 to 2 has increased the elastic domain of the plasticity part, so that the yield surface is not reached.
Thus, the softening from point 4 to 5 is only described by damage.
Only at point 5, the plasticity surface is reached and a subsequent increase of strain results in hardening of the plasticity part, which corresponds to an increase of the effective stress.
However, the nominal stress, shown in Fig.~\ref{fig:Cyclic}, decreases, since $\omega_{\rm c}$ increases.
The continuous slopes of parts 4-5 and 5-6 are obtained, since the additional factor $\beta_{\rm c}$ in (\ref{eq:kappaDC1}) is introduced.
A second reversal of the strain direction (point 6) changes the stress from compression to tension at point 7, which is again associated with a change of the stiffness.
The above response is very different from the one obtained with CDPM1 with only one damage parameter, which is also shown in Fig.~\ref{fig:Cyclic} by a dashed line. With CDPM1, the compressive response after point 3 is characterised by both a reduced stiffness and strength which would depend on the amount of damage accumulated in tension. For the case of damage equal to 1 in tension, both the strength and stiffness in compression would be zero, which is not realistic for the tension-compression transition in concrete.

\section{Mesh adjusted softening modulus} \label{sec:softeningModulus}
If the constitutive model described in the previous sections is straightaway used within the finite element method, the amount of dissipated energy might be strongly mesh-dependent.
This mesh-dependence is caused by localisation of deformations in mesh-size dependent zones. 
The finer the mesh, the less energy would be dissipated.
This is a well known limitation of constitutive laws with strain softening.
One way to overcome this mesh-dependence is to adjust the softening modulus with respect to the element size.
For the present model, this approach is applied for the tensile damage variable by replacing in the tensile damage law in (\ref{eq:damTenBilinear}) the strain thresholds $\varepsilon_{\rm f1}$ and $\varepsilon_{\rm f}$ with $w_{\rm f1}/h$ and $w_{\rm f}/h$, respectively.
Here, $w_{\rm f1}$ and $w_{\rm f}$ are displacement thresholds and $h$ is the finite element size.
Thus, with this approach the damage variables for bilinear softening are
\begin{equation}
\omega_{\rm t} = \left \{ \begin{array}{ll}
\dfrac{(E \kappa_{\rm dt} -f_{\rm t}) w_{\rm f1} - (\sigma_1 - f_{\rm t}) \kappa_{\rm dt1} h}{E \kappa_{\rm dt} w_{\rm f1} + (\sigma_{1} - f_{\rm t}) \kappa_{\rm dt2} h} & \mbox{if $0 < h \varepsilon_{\rm i} \leq w_{\rm f1} h$} \vspace{0.2cm}\\
\dfrac{E \kappa_{\rm dt} \left(w_{\rm f} - w_{\rm f1}\right) + \sigma_{\rm 1} \left(\kappa_{\rm dt1} h - w_{\rm f}\right)}{E \kappa_{\rm dt} \left(w_{\rm f} - w_{\rm f1}\right) - \sigma_{1} \kappa_{\rm dt2} h} & \mbox{if $w_{\rm f1} < h \varepsilon_{\rm i} \leq w_{\rm f}$} \vspace{0.2cm}\\
0  & \mbox{if $w_{\rm f} < h \varepsilon_{\rm i}$}\\
\end{array}
\right.
\end{equation}
These expressions are used when the constitutive model is compared to experimental results in the next section.
However, the evolution law for compressive damage is kept to be independent of the element size, as compressive failure is often accompanied by mesh-independent zones of localised displacements.

\section{ Implementation}
The present constitutive model has been implemented within the framework of the nonlinear finite element method, 
where the continuous loading process is replaced by incremental time steps. 
In each step the boundary value problem (global level) and the integration of the constitutive laws (local level) are solved.

For the boundary value problem on the global level, the usual incremental-iterative solution strategy is used, in the form of a modified Newton-Raphson iteration method. 
For the local problem, the updated values $\left(\cdot\right)^{\left(n+1\right)}$ of the stress and the internal variables at the end of the step are obtained by a fully implicit (backward Euler) integration of the rate form of the constitutive equations, starting from their known values $\left(\cdot\right)^{\left(n\right)}$ at the beginning of the step and applying the given strain increment $\Delta \boldsymbol{\varepsilon} = \boldsymbol{\varepsilon}^{\left(n+1\right)} - \boldsymbol{\varepsilon}^{\left(n\right)}$.
The integration scheme is divided into two sequential steps, corresponding to the plastic and damage parts of the model.
In the plastic part, the plastic strain $\boldsymbol{\varepsilon}_{\rm p}$ and the effective stress $\bar{\boldsymbol{\sigma}}$ at the end of the step  are determined. In the damage part, the damage variables $\omega_{\rm t}$ and $\omega_{\rm c}$, and the nominal stress $\boldsymbol{\sigma}$ at the end of the step are obtained.
The implementation strategy for the local problem, described in detail in \citet{GraJir06} for CDPM1, applies to the present model as well. To improve the robustness of the model, a subincrementation scheme is employed for the integration of the plasticity part.

\section{Comparison with experimental results} \label{sec:experiments}

In this section, the model response is compared to five groups of experiments reported in the literature.
For each group of experiments, the physical constants Young's modulus $E$, Poisson's ratio $\nu$, tensile strength $f_{\rm t}$, compressive strength $f_{\rm c}$ and tensile fracture energy $G_{\rm Ft}$ are adjusted to obtain a fit for the different types of concrete used in the experiments. 
The first four constants are model parameters. 
The last physical constant, $G_{\rm Ft}$, is directly related to model parameters.
For the bilinear softening law in section~\ref{sec:bilinear}, the tensile fracture energy is
\begin{equation}
G_{\rm Ft} = f_{\rm t} w_{\rm f1}/2 + \sigma_{1} w_{\rm f}/2
\end{equation} 
For $\sigma_1/f_{\rm t} = 0.3$ and $w_{\rm f1}/w_{\rm f} = 0.15$ (shown by \citet{JirZim98} to result in a good fit for concrete failure), the expression for the fracture energy reduces to $G_{\rm Ft} = f_{\rm t} w_{\rm f}/4.444$.
The compressive energy is $G_{\rm Fc} =  f_{\rm c} \varepsilon_{\rm fc} l_{c} A_{\rm s}$, where $l_{\rm c}$ is the length in which the compressive displacement are assumed to localise and $A_{\rm s}$ is the ductility measure in Sec.~2.3.4.
If no experimental results are available, the five constants can be determined using, for instance, the CEB-FIP Model Code \citep{CEB91}.

The other model parameters are set to their default values for all groups. 
The eccentricity constant $e$ that controls the shape of the deviatoric section is evaluated using the formula in \citet{JirBaz01}, p.~365:
\begin{equation} \label{eq:eccen}       
e = \frac{1+\epsilon}{2-\epsilon} \mbox{, where }\epsilon = \frac{f_{\rm t}}{f_{\rm bc}}\frac{f_{\rm bc}^2-f_{\rm c}^2}{f_{\rm c}^2-f_{\rm t}^2}
\end{equation} 
where $f_{\rm bc}$ is the strength in equibiaxial compression, which is estimated as $f_{\rm bc} = 1.16 f_{\rm c}$ according to the experimental results reported in \citet{Kupfer69}.
Parameter $q_{h0}$ is the dimensionless ratio $q_{h0}=\bar{f}_{c0}/f_{\rm c}$, where $f_{\rm c0}$ is the compressive stress at which the initial yield limit is reached in the plasticity model for uniaxial compression. Its default value is $q_{h0} = 0.3$.
For the hardening modulus the default value is $H_{\rm p} = 0.01$.
Furthermore, the default value of the parameter of the flow rule is chosen as  $D_{\rm f} = 0.85$, which yields a good agreement with experimental results in uniaxial compression.
The determination of parameters $A_{\rm h}$, $B_{\rm h}$, $C_{\rm h}$ and $D_{\rm h}$ that influence the hardening ductility measure is more difficult. The effective stress varies within the hardening regime, even for monotonic loading, so that the ratio of axial and lateral plastic strain rate is not constant. Thus, an exact relation of all four model parameters to measurable material properties cannot be constructed. 
In \citet{GraJir06}, it has been shown that a reasonable response is obtained with parameters $A_{\rm h} = 0.08$, $B_{\rm h} = 0.003$, $C_{\rm h} = 2$ and $D_{\rm h} = 1 \times 10^{-6}$. 
These values were also used in the present study.
Furthermore, the element size $h$ in the damage laws in Section~\ref{sec:softeningModulus} was chosen as $h = 0.1$~m.

\begin{figure}[htb!]
\begin{center}
\includegraphics[width=10cm]{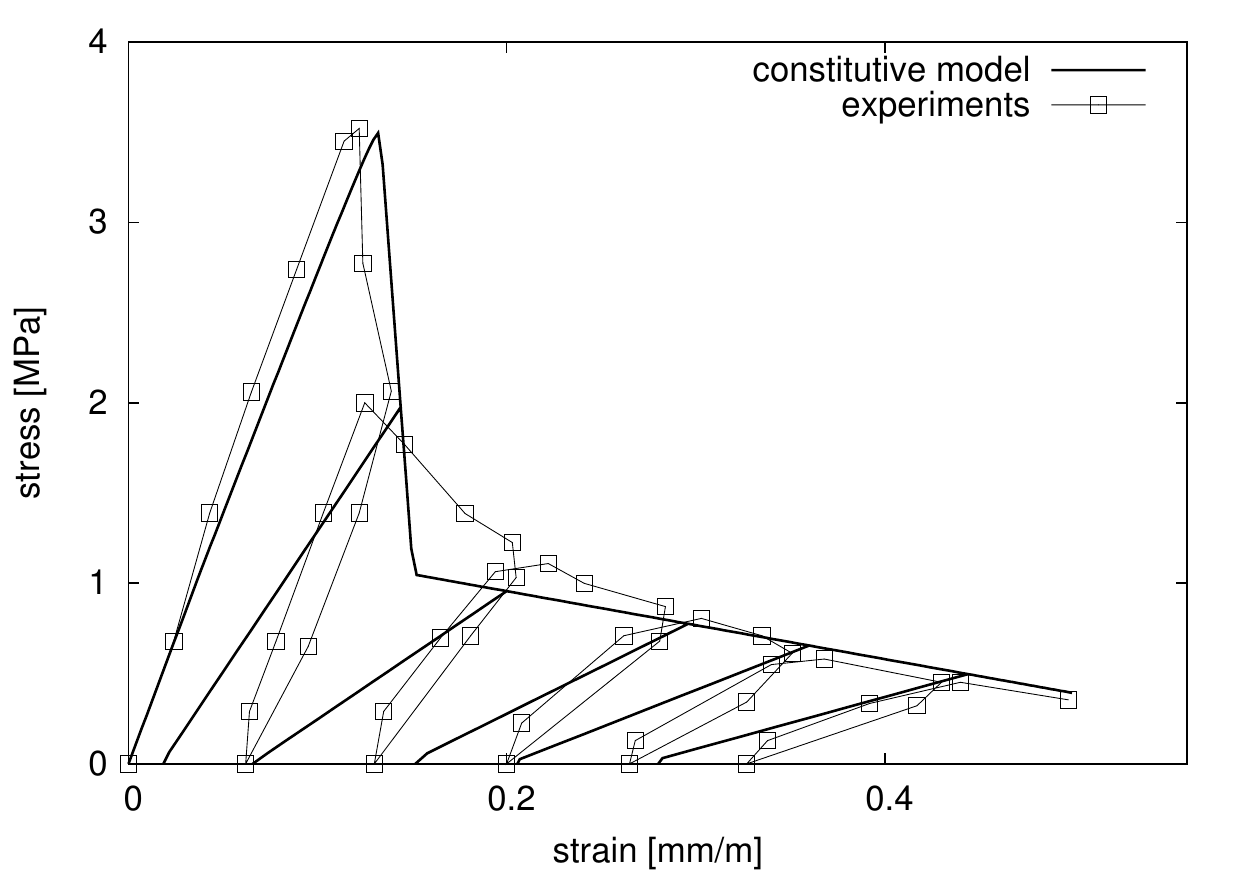}
\end{center}
\caption{Uniaxial tension: Model response compared to experimental results in \protect \citet{GopSha85}.}
\label{fig:GopSha85}
\end{figure}

The first analysis is a uniaxial tensile setup with unloading. 
The model response is compared to the experimental results reported in~\citet{GopSha85} (Fig.~\ref{fig:GopSha85}).
The relevant model parameters for this experiment are $E = 28$~GPa, $\nu = 0.2$, $f_{\rm c} = 40$~MPa, $f_{\rm t} = 3.5$~MPa, $G_{\rm Ft} = 55$~J/m$^2$. 

\begin{figure}[htb!]
\begin{center}
\includegraphics[width=10cm]{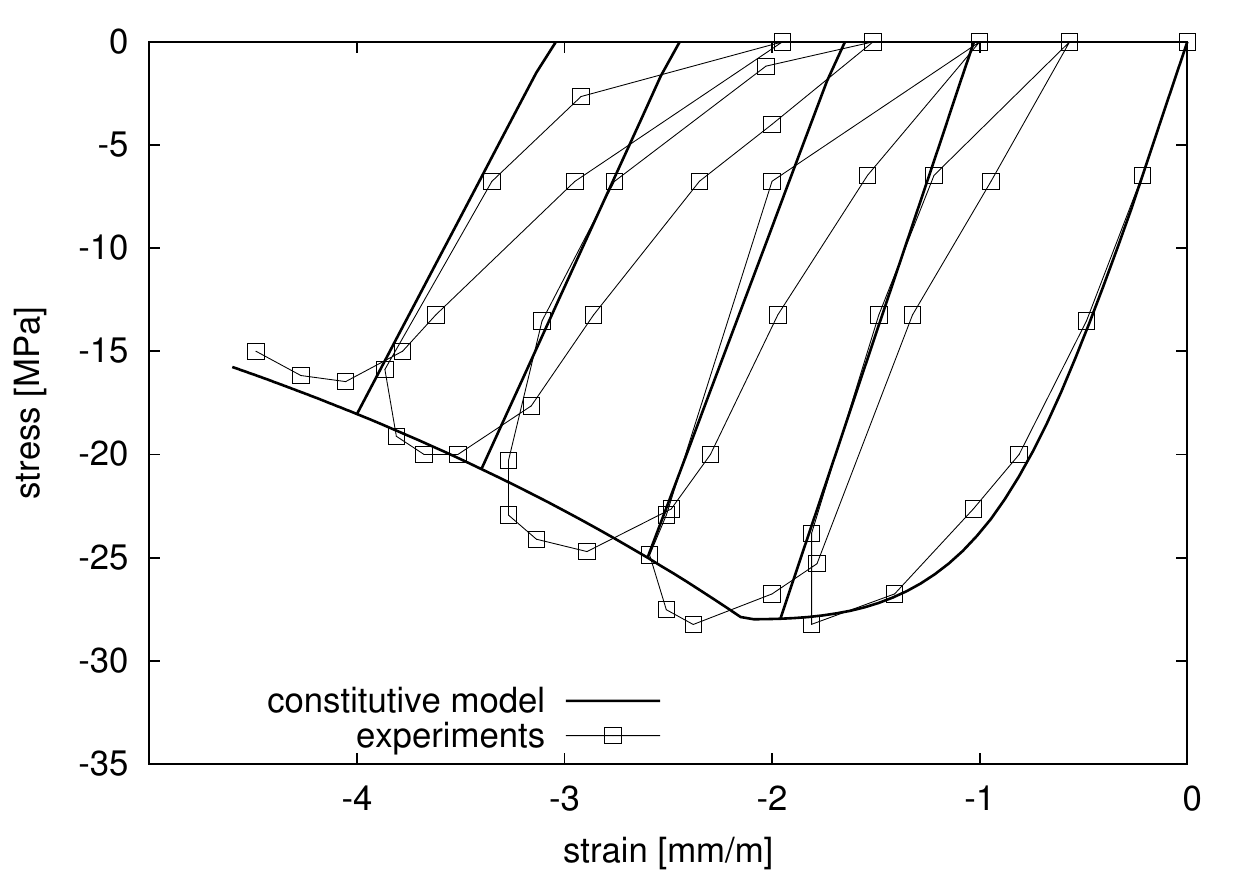}
\end{center}
\caption{Uniaxial compression: Model response compared to experimental results reported in \protect \citet{KarJir69}.}
\label{fig:KarJir69}
\end{figure}

The next example is an uniaxial compression test with unloading, for which the model response is compared to experimental results reported in~\citet{KarJir69} (Fig.~\ref{fig:KarJir69}).
The model parameters are $E = 30$~GPa, $\nu = 0.2$, $f_{\rm c} = 28$~MPa, $f_{\rm t} = 2.8$~MPa.
Furthermore, the model constants for compression are $A_{\rm s} = 5$ and $\varepsilon_{\rm fc} = 0.0001$.
The value of the tensile fracture energy $G_{\rm Ft}$ does not influence the model response in compression, which also applies to all other compression tests considered in the following paragraphs. Therefore, only the compressive fracture energy is stated.

\begin{figure}[htb!]
\begin{center}
\includegraphics[width=10cm]{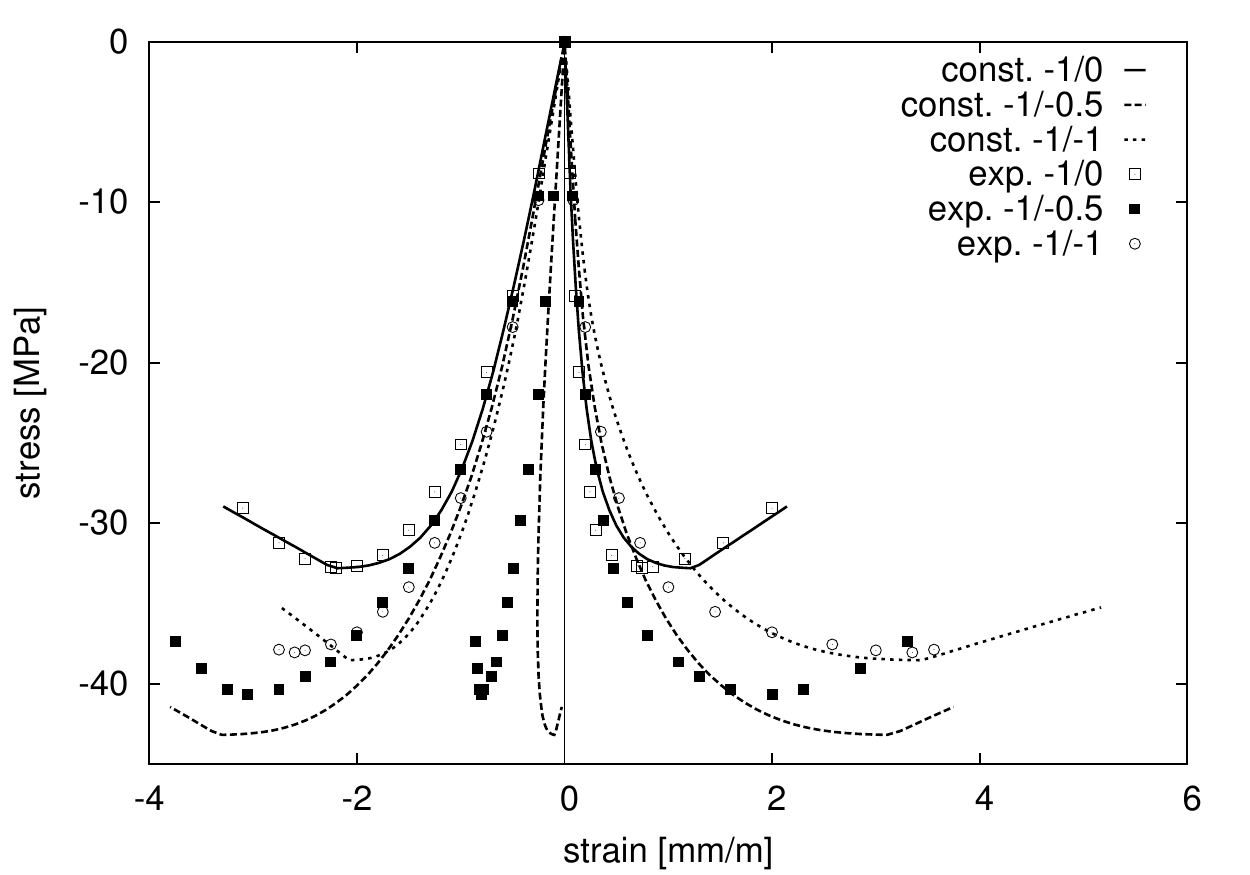}
\end{center}
\caption{Uniaxial and biaxial compression: Model response compared to experimental results reported in \protect \citet{Kupfer69}.}
\label{fig:Kupfer69}
\end{figure}

Next, the model is compared to uniaxial and biaxial compression tests reported in \citet{Kupfer69}.
For these experiments, the model parameters are set to $E = 32$~GPa, $\nu = 0.2$, $f_{\rm c} = 32.8$~MPa, $f_{\rm t} = 3.3$~MPa.
Furthermore, the model constants for compression are $A_{\rm s} = 1.5$ and $\varepsilon_{\rm fc} = 0.0001$.
The comparison with experimental results is shown in Fig.~\ref{fig:Kupfer69} for uniaxial, equibiaxial and biaxial compression.
For the biaxial compression case, the stress ratio of the two compressive stress components is $\sigma_1/\sigma_2 = -1/-0.5$.

\begin{figure}[htb!]
\begin{center}
\includegraphics[width=10cm]{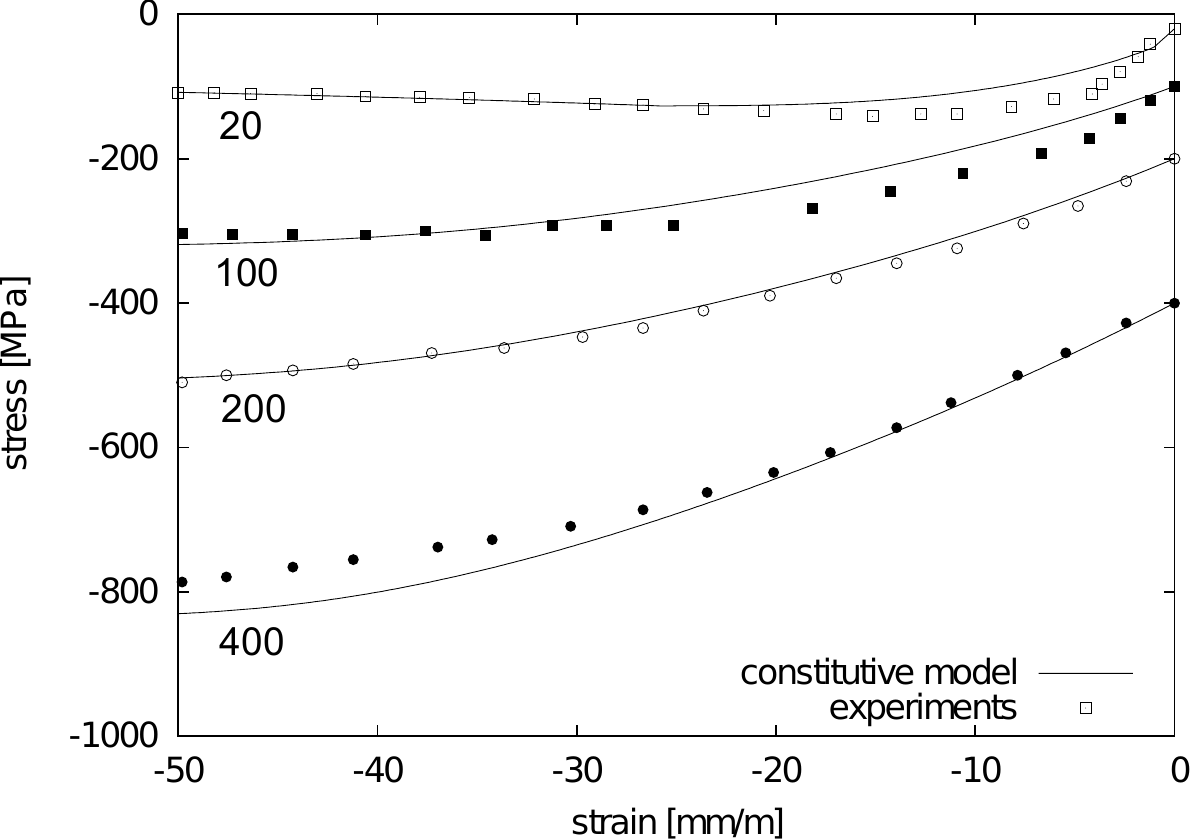}
\end{center}
\caption{Confined compression: Model response compared to experiments used in \protect \citet{CanBaz00}.}
\label{fig:WES1994Tri}
\end{figure}

Furthermore, the performance of the model is evaluated for triaxial tests reported in \citet{CanBaz00}. 
The material parameters for this test are $E = 25$~GPa, $\nu=0.2$,  $f_{\rm c} = 45.7$~MPa, $f_{\rm t} = 4.57$~MPa.
Furthermore, the model constants for compression are $A_{\rm s} = 15$ and $\varepsilon_{\rm fc} = 0.0001$.  
The model response is compared to experimental results presented in Figs.~\ref{fig:WES1994Tri}.

Next, the model response in triaxial compression is compared to the experimental results reported in \citet{ImrPan96}.
The material parameters for this test are $E = 30$~GPa, $\nu=0.2$,  $f_{\rm c} = 47.4$~MPa, $f_{\rm t} = 4.74$~MPa.
Furthermore, the model constants for compression are $A_{\rm s} = 15$ and $\varepsilon_{\rm fc} = 0.0001$.  
\begin{figure}[htb!]
\begin{center}
\includegraphics[width=10cm]{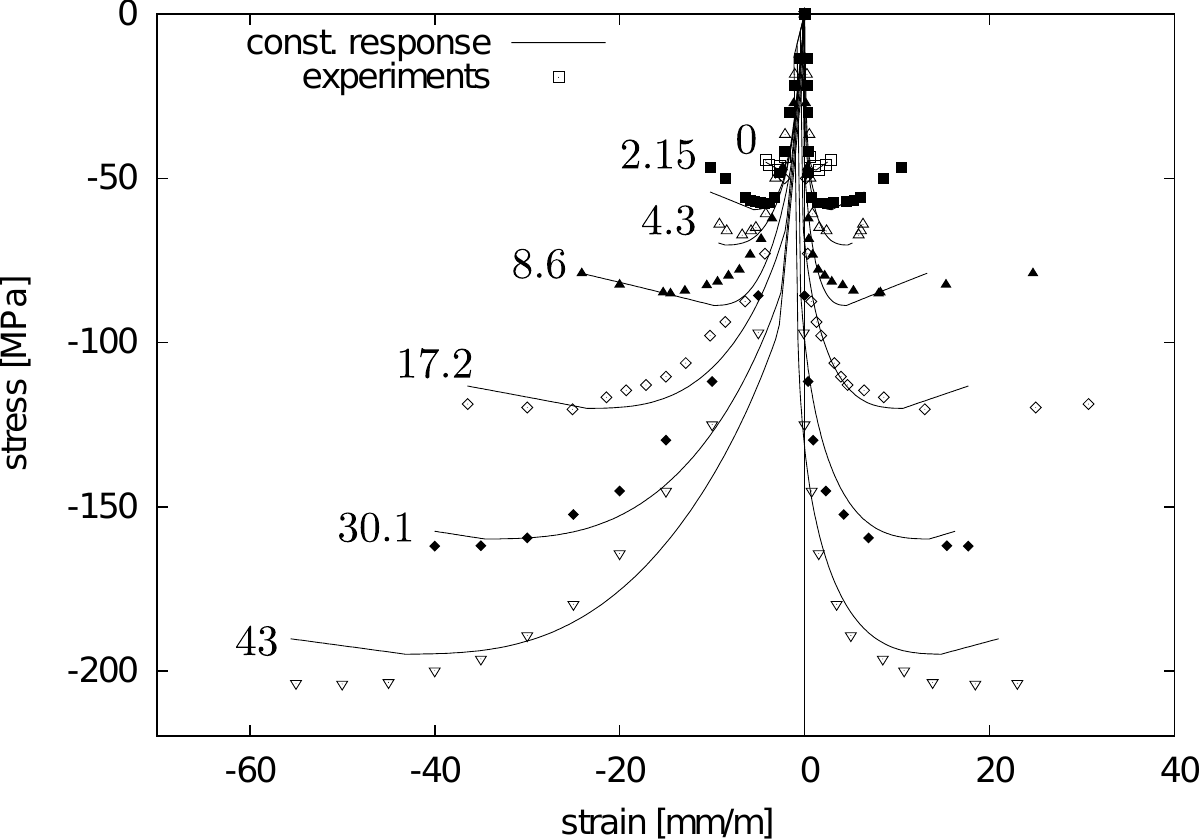}
\end{center}
\caption{Confined compression: Model response compared to experiments reported in \protect \citet{ImrPan96}.}
\label{fig:ImrPan96}
\end{figure}

Finally, the model response in hydrostatic compression is compared to the experimental results reported in \citet{CanBaz00}.
The material parameters are the same as for the triaxial test shown in Fig.~\ref{fig:WES1994Tri}.
\begin{figure}[htb!]
\begin{center}
\includegraphics[width=10cm]{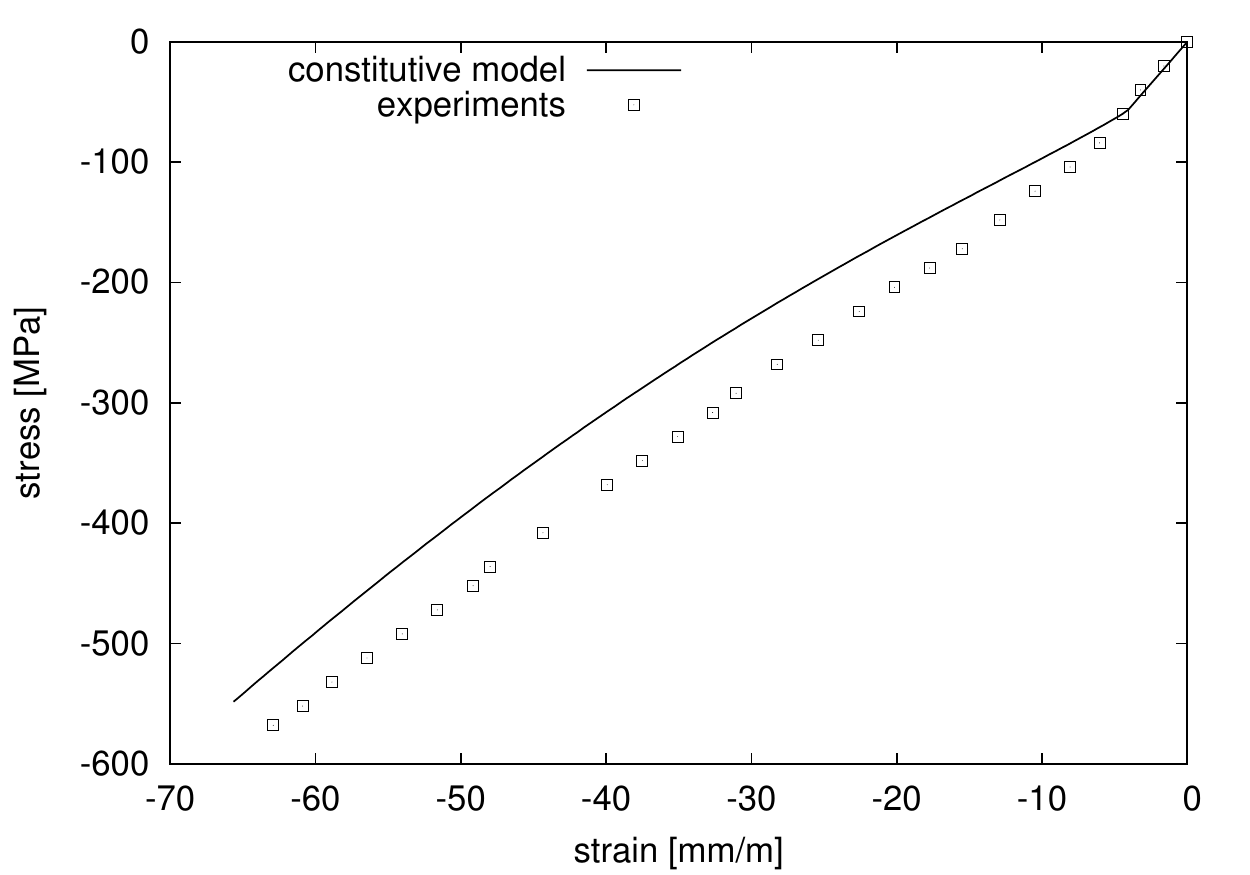}
\end{center}
\caption{Hydrostatic compression: Model response compared to experiments reported in \protect \citet{CanBaz00}.}
\label{fig:WES1994Hydro}
\end{figure}

Overall, the agreement of the model response with the experimental results is very good. 
The model is able to represent the strength of concrete in tension and multiaxial compression. 
In addition, the strains at maximum stress in tension and compression agree well with the experimental results.
The bilinear stress-crack opening curve that was used results in a good approximation of the softening curve in uniaxial tension and compression.
With the above comparisons, it is demonstrated that CDPM2, provides, very similar to CDPM1, a very good agreement with experimental results.

\section{Structural analysis} 

The performance of the proposed constitutive model is further evaluated by structural analysis of three fracture tests.
The main objective of this part of the study is to demonstrate that the structural response obtained with the model is mesh-independent.
This is achieved by adjusting the softening modulus with respect to the element size (section~\ref{sec:softeningModulus}).

\subsection{Three point bending test}
\begin{figure}[htb!]
\begin{center}
\includegraphics[width=10cm]{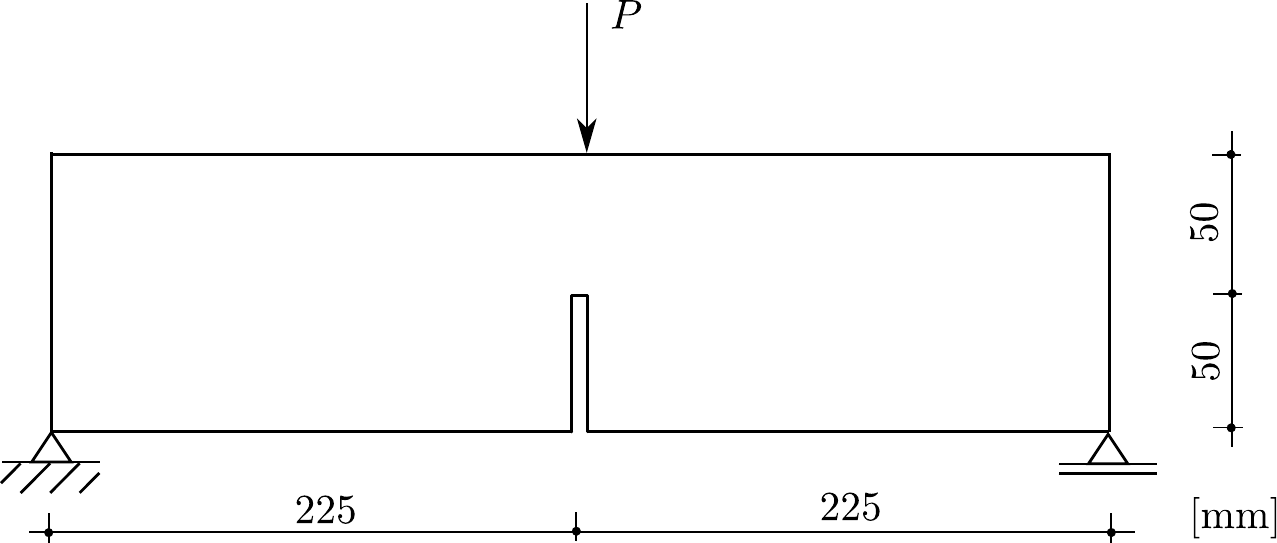}
\end{center}
\caption{Three point bending test: Geometry and loading setup. The out-of-plane thickness is $0.1$~m. The notch thickness is $5$~mm.}
\label{fig:3pbtGeometry}
\end{figure}

\begin{figure}[htb!]
\begin{center}
\includegraphics[width=10cm]{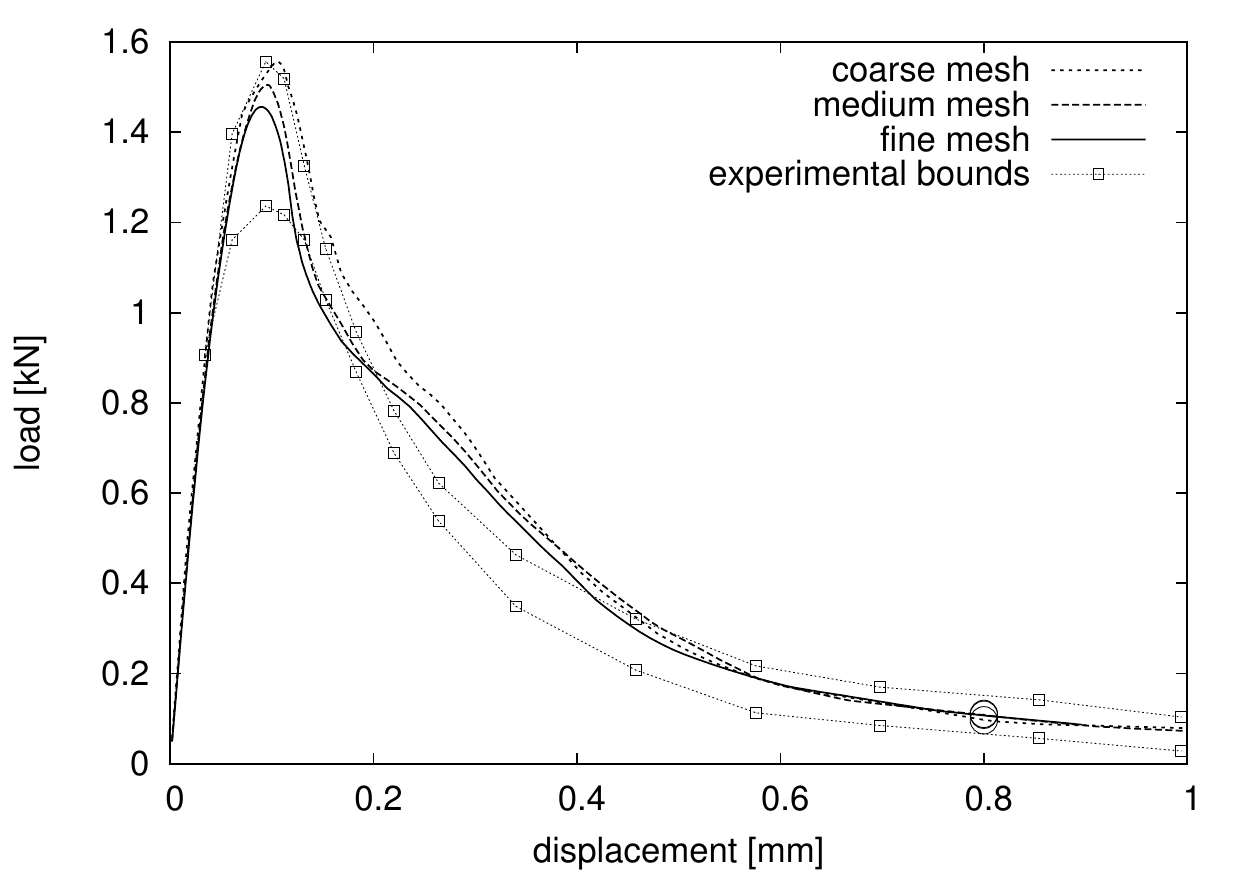}
\end{center}
\caption{Load-CMOD curves of analyses with three mesh sizes compared to the experimental bounds reported in \protect \citet{KorRei83}.}
\label{fig:3pbtLoad}
\end{figure}

\begin{figure}[htb!]
\begin{center}
\includegraphics[width=12cm]{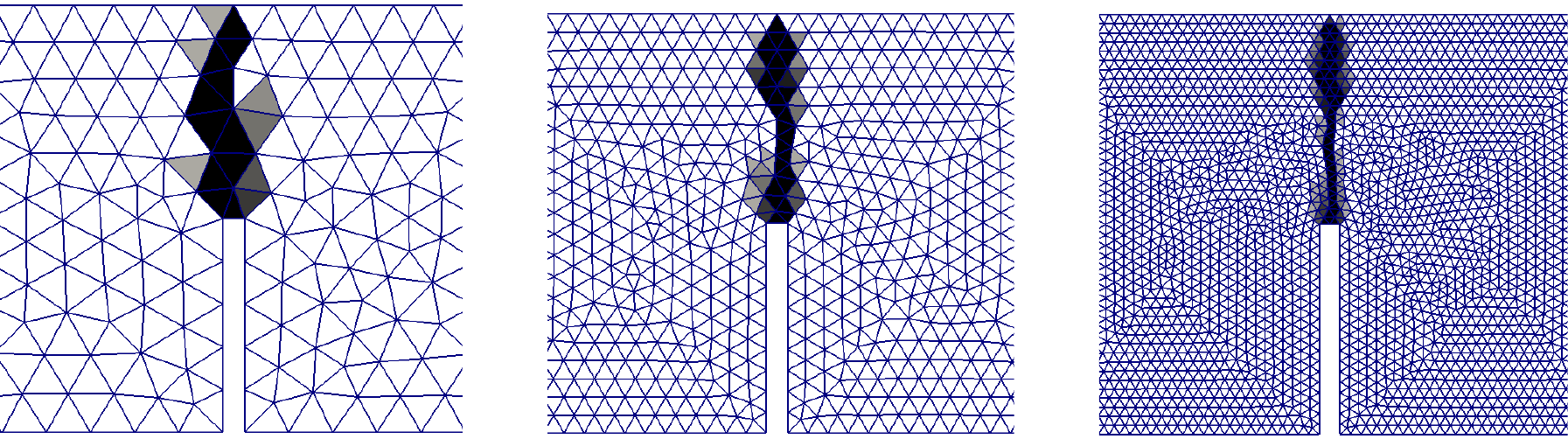}
\end{center}
\caption{Tensile damage patterns for the coarse, medium and fine mesh for the three point bending test. Black indicates a tensile damage variable of 1.}
\label{fig:3pbtDamage}
\end{figure}

The first structural example is a three-point bending test of a single-edge notched beam reported by \citet{KorRei83}. 
The experiment is modelled by triangular plane strain finite elements with three mesh sizes. 
The geometry and loading set up is shown in Fig.~\ref{fig:3pbtGeometry}.
The input parameters are chosen as $E=20$~GPa, $\nu=0.2$, $f_{\rm t}=2.4$~MPa, $G_{\rm ft}=100$~N/m, $f_{\rm c}=24$~MPa \citep{GraJir06a}. All other parameters are set to their default values described in section~\ref{sec:experiments}.
For this type of analysis, local stress-strain relations with strain softening are known to result in mesh-dependent load-displacement curves.
The capability of the adjustment of the softening modulus approach presented in section~\ref{sec:softeningModulus} to overcome this mesh-dependence is assessed with this test.
The global response in the form of load-Crack Mouth Opening Displacement (CMOD) is shown in Fig.~\ref{fig:3pbtLoad}.
The local response in the form of tensile damage patterns at loading stages marked in Fig.~\ref{fig:3pbtLoad} for the three meshes is shown in Fig.~\ref{fig:3pbtDamage}.

Overall, the load-CMOD curves in Fig.~\ref{fig:3pbtLoad} are in good agreement with the experimental results and almost mesh independent. 
On the other hand, the damage zones in Fig.~\ref{fig:3pbtDamage} depend on the mesh size.

\subsection{Four point shear test}
\begin{figure}[htb!]
\begin{center}
\includegraphics[width=10cm]{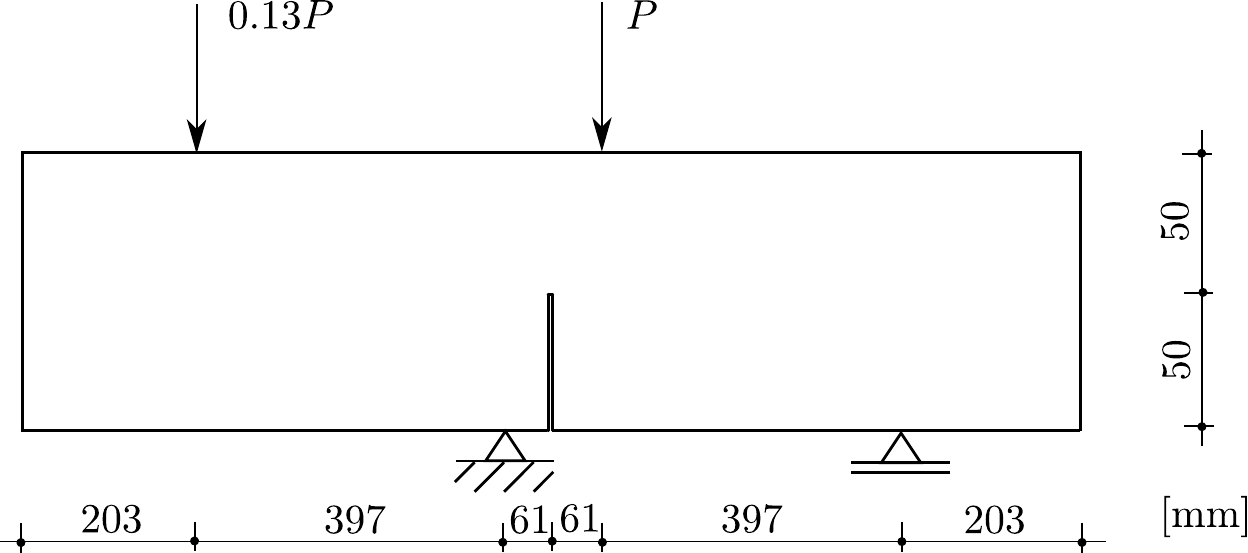}
\end{center}
\caption{Four point shear test: Geometry and loading setup. The out-of-plane thickness is $0.15$~m. A zero notch thickness is assumed.}
\label{fig:arreaGeometry}
\end{figure}

\begin{figure}[htb!]
\begin{center}
\includegraphics[width=10cm]{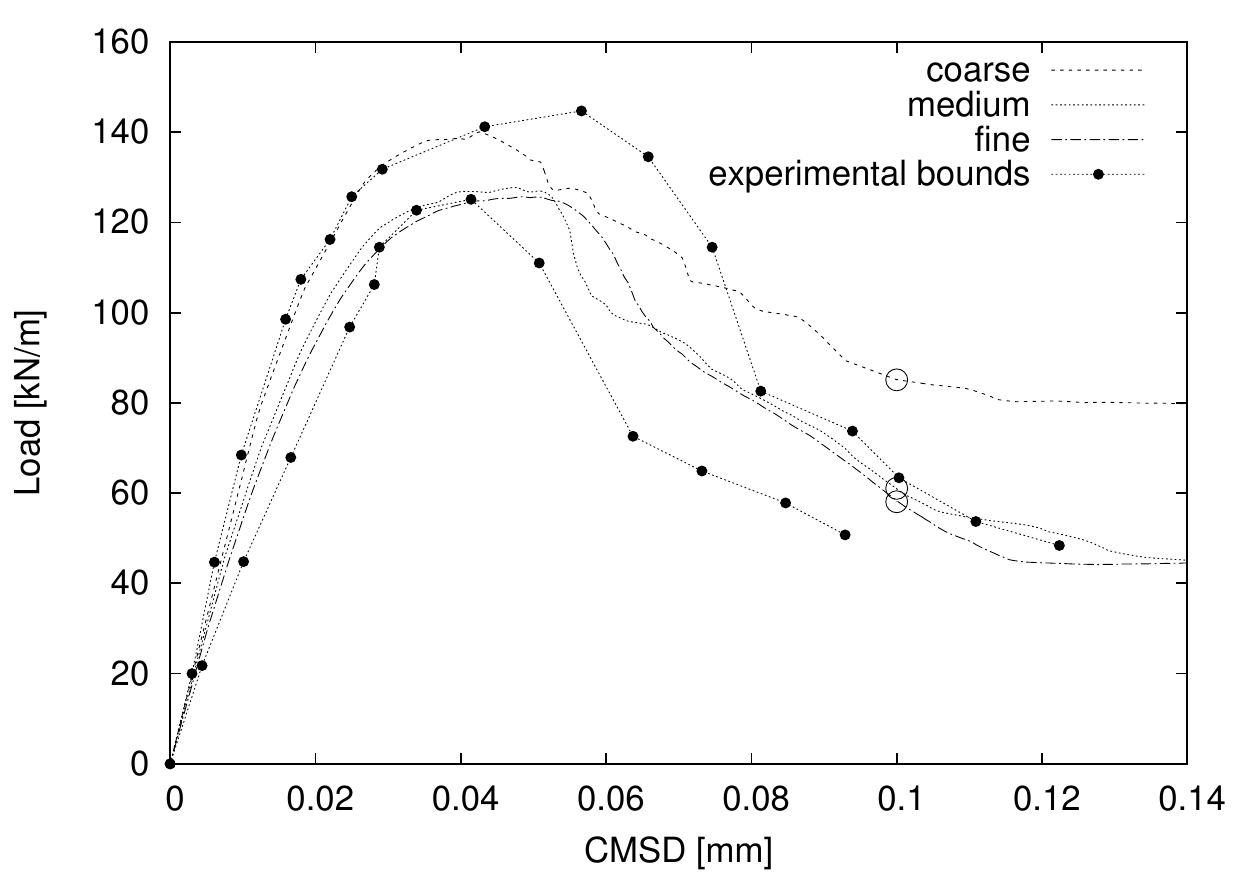}
\end{center}
\caption{Load-CMSD curves of analyses with three mesh sizes compared to the experimental bounds reported in \protect \citet{ArrIng82}.}
\label{fig:arreaLoad}
\end{figure}

\begin{figure}[htb!]
\begin{center}
\includegraphics[width=8cm]{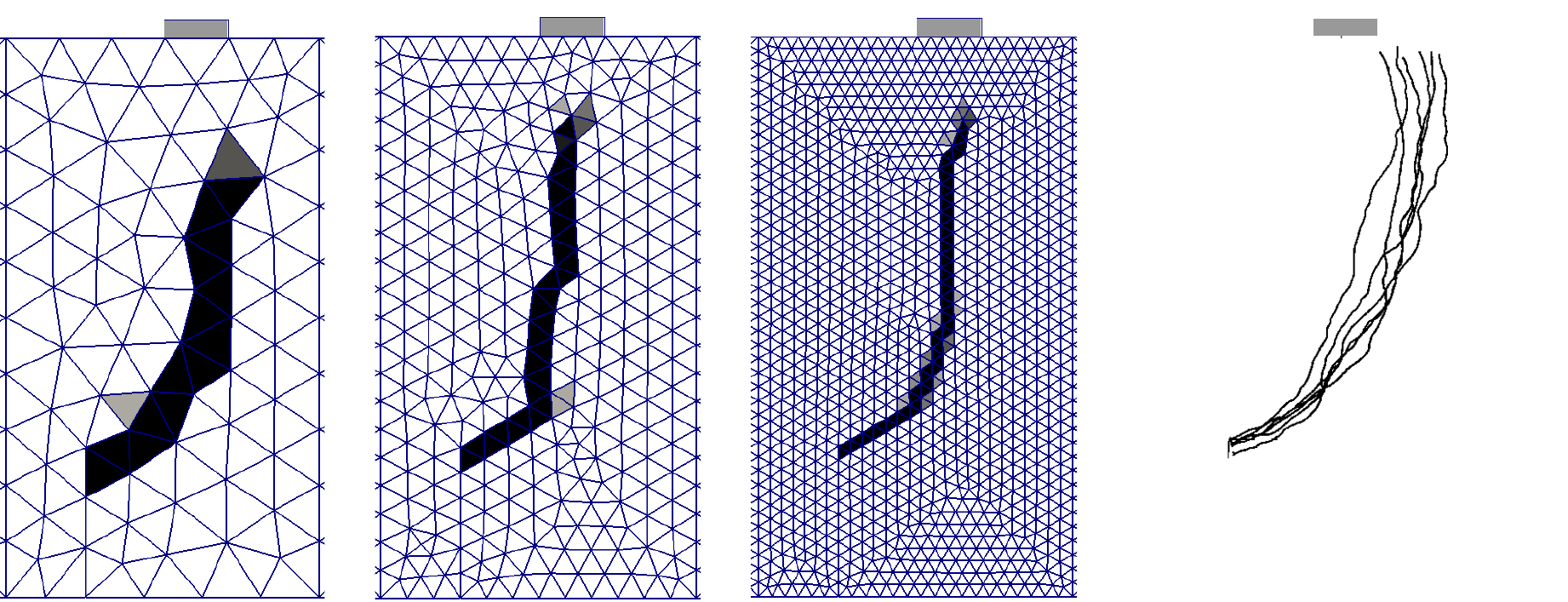}
\end{center}
\caption{Four point shear test: Tensile damage patterns for the coarse, medium and fine mesh compared to the experimental crack patterns reported in \protect \citet{ArrIng82}. Black indicates a tensile damage variable of 1.}
\label{fig:arreaDamage}
\end{figure}

The second structural example is a four point shear test of a single-edge notched beam reported in \citet{ArrIng82}. Again, the experiment is modelled by triangular plane strain finite elements with three different mesh sizes.
The geometry and loading setup are shown in Fig.~\ref{fig:arreaGeometry}. 
The input parameters are chosen as $E=30$~GPa, $\nu = 0.18$, $f_{\rm t} = 3.5$~MPa, $G_{\rm ft} = 140$~N/m, $f_{\rm c} = 35$~MPa \citep{JirGra08}. 
All other parameters are set to their default values described in section~\ref{sec:experiments}.
The global responses of analyses and experimental results are compared in the form of load-Crack Mouth Sliding Displacement (CMSD) curves in Fig.~\ref{fig:arreaLoad}.
Furthermore, the damage patterns for the three meshes at loading stages marked in Fig.~\ref{fig:arreaLoad} are compared to the experimental crack patterns in Fig.~\ref{fig:arreaDamage}.

The load-CMSD curves obtained with the three meshes are in good agreement with the experimental results.
The coarse mesh overestimates the load levels obtained with the medium and fine mesh.
However, the two finer meshes are in good agreement.
Again, the width of the damaged zone depends on the element size. 
Furthermore, the damage zones are influenced by the mesh orientation. 
In particular, for the fine mesh the damage zone follows the regular element arrangement, so that the crack is less curved than reported in the experiments.
This is a well known behaviour of models using the adjustement of the softening modulus approach, which has been studied in more detail in \citet{JirGra08,GraRem07}. 

\subsection{Eccentric compression test}
\begin{figure}
\begin{center}
\begin{tabular}{cc}
\includegraphics[height=7cm]{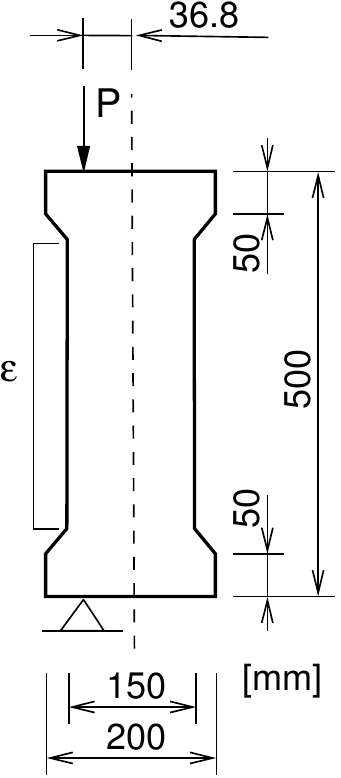} & \hskip 0.5cm \includegraphics[height=6cm]{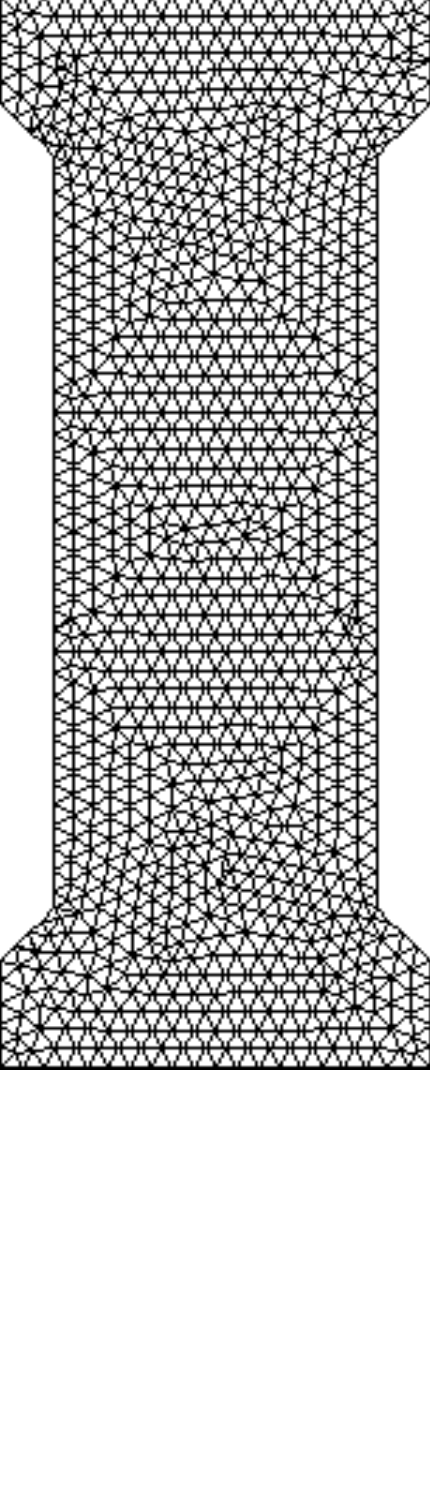}\\
(a) & (b)
\end{tabular} 
\end{center}
\caption{a) Geometry and loading setup of the eccentric compression test. b) The coarse finite element mesh.} 
\label{fig:DebernardiGeometry}
\end{figure}

The third structural example studies the failure of a concrete prism subjected to eccentric compression, tested by \cite{DebTal01}. The geometry and loading setup are shown in Figure~\ref{fig:DebernardiGeometry}a. The specimen with a relatively great eccentricity of $36.8$ mm is modeled by a thin layer of linear 3D elements to reduce the computational time compared to a full 3D analysis. Three different mesh sizes with element lengths of 7.5, 5 and 2.5 mm were chosen (see Figure~\ref{fig:DebernardiGeometry}b for the coarse mesh).

\begin{figure}
\begin{center}
\includegraphics[width=10cm]{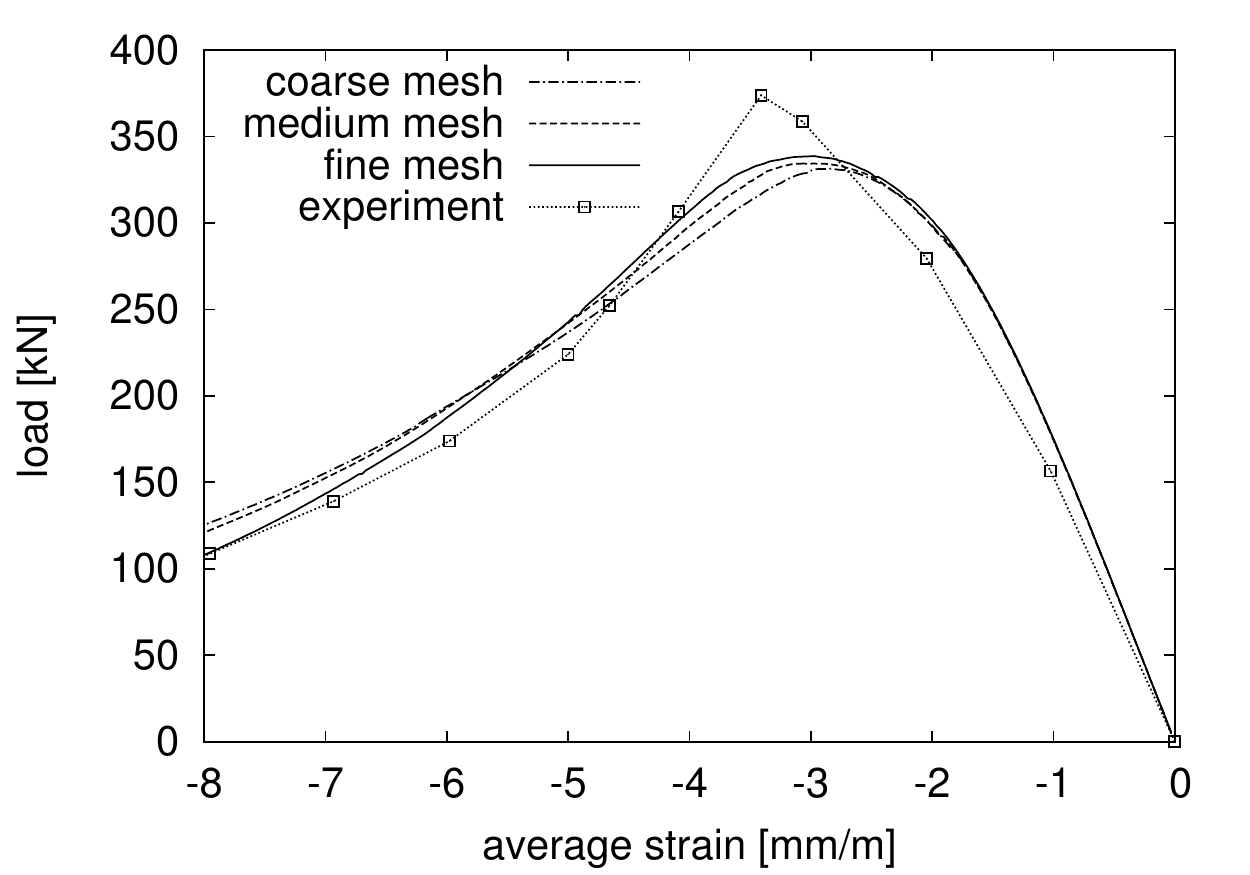}
\end{center}
\caption{Comparison of the analysis of the eccentric compression test with the experiment.} 
\label{fig:debernardiExpComp}
\end{figure}

\begin{figure}
\begin{center}
\begin{tabular}{ccc}
\includegraphics[width=2cm]{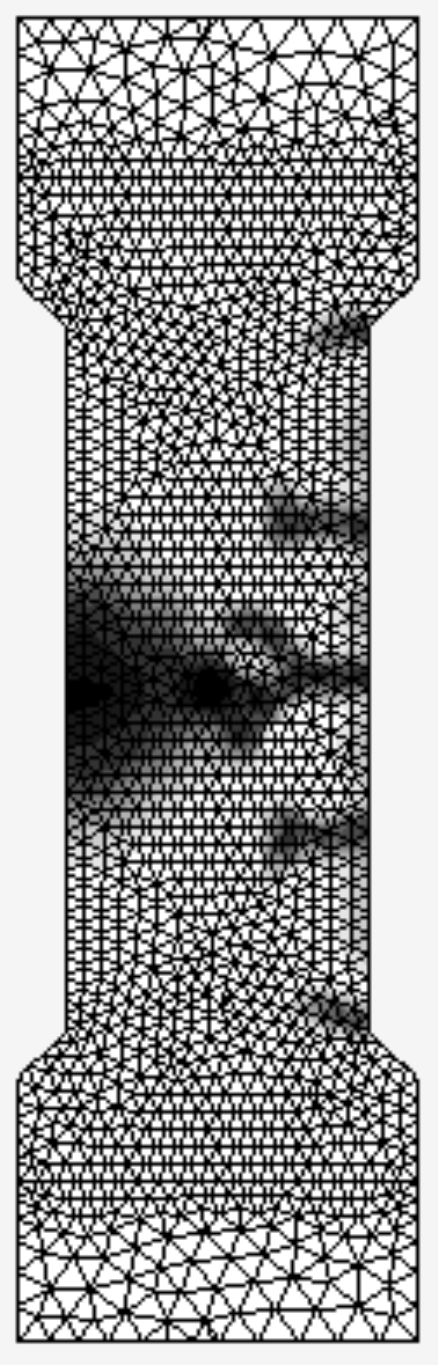} & \includegraphics[width=2cm]{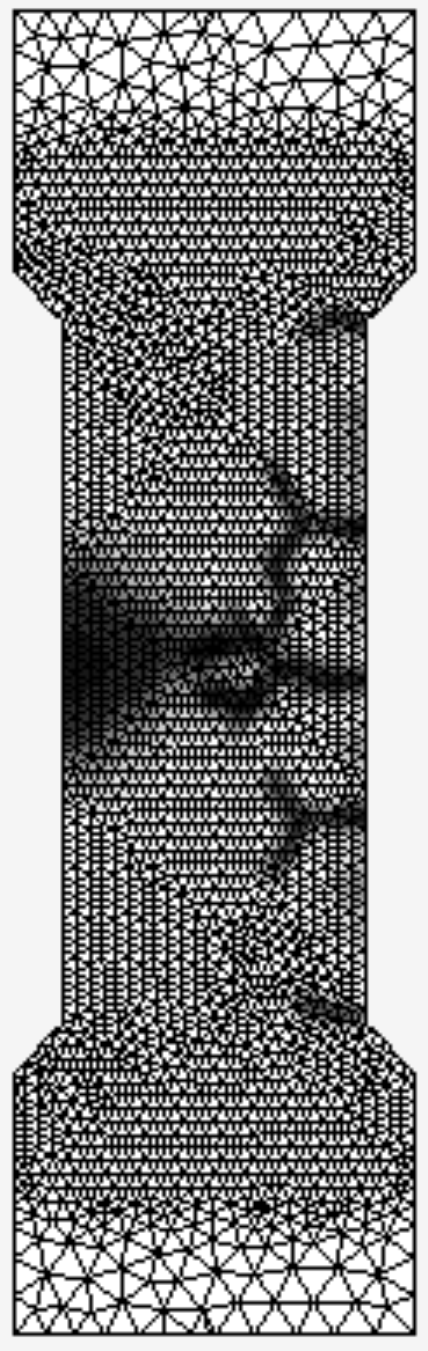} & \includegraphics[width=2.05cm]{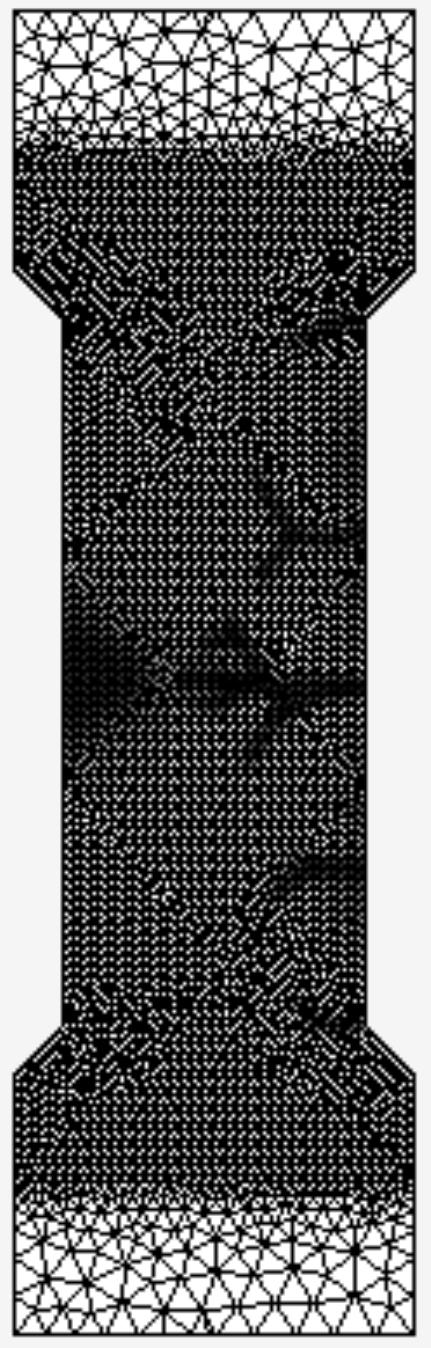}\\
(a) & (b) & (c)
\end{tabular} 
\end{center}
\caption{Contour plots of the damage variable for the a) coarse, b) medium and c) fine mesh of the eccentric compression test.} 
\label{fig:debernardiContour}
\end{figure}

The model parameters were set to $E=30$~GPa, $\nu=0.2$, $f_{\rm t} = 4$~MPa, $f_{\rm c} = 46$~MPa, $G_{\rm Ft}=100$~N/m, $A_{\rm s} = 10$ and $\varepsilon_{\rm fc} = 0.0001$.
The model response in terms of the overall load versus the mean compressive strain of the compressed side obtained on the fine mesh is compared to the experimental result in Figure~\ref{fig:debernardiExpComp}.
The load capacity and the strain at peak are underestimated by the model. 
The overall behaviour, however, is captured well.
The comparison of the load-compressive strain relations for the analyses on meshes of different sizes indicates that the description of this type of compressive failure is nearly mesh-independent.
The evolution of the damage zone for the analysis on the coarse mesh is depicted in Figure~\ref{fig:debernardiContour} for the final stage of the analyses in Figure~\ref{fig:debernardiExpComp}.
On the tensile side several zones of localized damage form, whereas the failure on the compressive side is described by a diffuse damage zone.

\section{Conclusions}
The present damage plasticity model CDPM2, which combines a stress-based plasticity part with a strain based damage mechanics model, is based on an enhancement of an already exisiting damage-plasticity model called CDPM1 (\citet{GraJir06}). Based on the work presented in this manuscript, the following conclusions can be drawn on the improvements that this constitutive model provides:\\
\begin{enumerate}
\item The model is able to describe realistically the transition from tensile to compressive failure. This is achieved by the introduction of two separate isotropic damage variables for tension and compression.\\
\item The model is able to reproduce stress inelastic strain relations with varying ratios of reversible and irreversible strain components. The ratio can be controlled by the hardening modulus of the plasticity part.
\item The model gives meshindependent load-displacement curves for both tensile and compressive failure.
\end{enumerate}

In addition, the model response is in good agreement with experimental results for a wide range of loading from uniaxial tension to confined compression.

\section*{Acknowledgements}
This work have been performed partially within the project ``Dynamic behaviour of reinforced concrete structures subjected to blast and fragment impacts'', which in turn is financially sponsored by MSB - the Swedish Civil Contigencies Agency.
The first author would also like to thank Prof. Bo\v{r}ek Patz\'ak of the Czech Technical University for kind assistance with his finite element package OOFEM \citep{Pat99,PatBit01}.
In addition, the first and second authors acknowledge funding received from the UKs Engineering and Physical Sciences Research Council (EPSRC) under grant EP/I036427/1.

\bibliographystyle{model1b-num-names}
\bibliography{general}

\end{document}


%% file: report.bbl
\begin{thebibliography}{45}
\expandafter\ifx\csname natexlab\endcsname\relax\def\natexlab#1{#1}\fi
\providecommand{\url}[1]{\texttt{#1}}
\providecommand{\href}[2]{#2}
\providecommand{\path}[1]{#1}
\providecommand{\DOIprefix}{doi:}
\providecommand{\ArXivprefix}{arXiv:}
\providecommand{\URLprefix}{URL: }
\providecommand{\Pubmedprefix}{pmid:}
\providecommand{\doi}[1]{\href{http://dx.doi.org/#1}{\path{#1}}}
\providecommand{\Pubmed}[1]{\href{pmid:#1}{\path{#1}}}
\providecommand{\bibinfo}[2]{#2}
\ifx\xfnm\relax \def\xfnm[#1]{\unskip,\space#1}\fi
\bibitem[{Arrea and Ingraffea(1982)}]{ArrIng82}
\bibinfo{author}{M.~Arrea}, \bibinfo{author}{A.R. Ingraffea},
  \bibinfo{title}{Mixed-mode crack propagation in mortar and concrete},
  \bibinfo{type}{Department of Structural Engineering} \bibinfo{number}{81-83},
  Cornell University, \bibinfo{address}{Ithaca, NY}, \bibinfo{year}{1982}.
\bibitem[{Caner and Ba\v{z}ant(2000)}]{CanBaz00}
\bibinfo{author}{F.C. Caner}, \bibinfo{author}{Z.P. Ba\v{z}ant},
  \bibinfo{title}{Microplane model {M4} for concrete: {II.~Algorithm} and
  calibration}, \bibinfo{journal}{\JEM} \bibinfo{volume}{126}
  (\bibinfo{year}{2000}) \bibinfo{pages}{954--961}.
\bibitem[{Carol et~al.(2001)Carol, Rizzi and Willam}]{CarRizWil01a}
\bibinfo{author}{I.~Carol}, \bibinfo{author}{E.~Rizzi}, \bibinfo{author}{K.J.
  Willam}, \bibinfo{title}{On the formulation of anisotropic elastic
  degradation. {I. T}heory based on a pseudo-logarithmic damage tensor rate},
  \bibinfo{journal}{\IJSS} \bibinfo{volume}{38} (\bibinfo{year}{2001})
  \bibinfo{pages}{491--518}.
\bibitem[{CEB(1991)}]{CEB91}
\bibinfo{author}{CEB}, \bibinfo{title}{{CEB-FIP Model Code 1990, Design Code}},
  \bibinfo{publisher}{Thomas Telford}, \bibinfo{address}{London},
  \bibinfo{year}{1991}.
\bibitem[{Debernardi and Taliano(2001)}]{DebTal01}
\bibinfo{author}{P.G. Debernardi}, \bibinfo{author}{M.~Taliano},
  \bibinfo{title}{Softening behaviour of concrete prisms under eccentric
  compressive forces}, \bibinfo{journal}{\MCR} \bibinfo{volume}{53}
  (\bibinfo{year}{2001}) \bibinfo{pages}{239--249}.
\bibitem[{Etse and Willam(1994)}]{Etse94}
\bibinfo{author}{G.~Etse}, \bibinfo{author}{K.J. Willam}, \bibinfo{title}{A
  fracture-energy based constitutive formulation for inelastic behavior of
  plain concrete}, \bibinfo{journal}{\JEM} \bibinfo{volume}{120}
  (\bibinfo{year}{1994}) \bibinfo{pages}{1983--2011}.
\bibitem[{Fichant et~al.(1999)Fichant, Borderie and
  Pijaudier-Cabot}]{FicBorPij99}
\bibinfo{author}{S.~Fichant}, \bibinfo{author}{C.L. Borderie},
  \bibinfo{author}{G.~Pijaudier-Cabot}, \bibinfo{title}{Isotropic and
  anisotropic descriptions of damage in concrete structures},
  \bibinfo{journal}{Mechanics of Cohesive-Frictional Materials}
  \bibinfo{volume}{4} (\bibinfo{year}{1999}) \bibinfo{pages}{339--359}.
\bibitem[{Folino and Etse(2012)}]{FolEts12}
\bibinfo{author}{P.~Folino}, \bibinfo{author}{G.~Etse},
  \bibinfo{title}{Performance dependent model for normal and high strength
  concretes}, \bibinfo{journal}{International Journal of Solids and Structures}
  \bibinfo{volume}{49} (\bibinfo{year}{2012}) \bibinfo{pages}{701--719}.
\bibitem[{Gopalaratnam and Shah(1985)}]{GopSha85}
\bibinfo{author}{V.S. Gopalaratnam}, \bibinfo{author}{S.P. Shah},
  \bibinfo{title}{Softening response of plain concrete in direct tension},
  \bibinfo{journal}{ACI Journal Proceedings} \bibinfo{volume}{82}
  (\bibinfo{year}{1985}).
\bibitem[{Grassl(2009)}]{Gra09b}
\bibinfo{author}{P.~Grassl}, \bibinfo{title}{On a damage-plasticity approach to
  model concrete failure}, \bibinfo{journal}{Proceedings of the ICE -
  Engineering and Computational Mechanics} \bibinfo{volume}{162}
  (\bibinfo{year}{2009}) \bibinfo{pages}{221--231}.
\bibitem[{Grassl and Jir{\'a}sek(2006)}]{GraJir06}
\bibinfo{author}{P.~Grassl}, \bibinfo{author}{M.~Jir{\'a}sek},
  \bibinfo{title}{{Damage-plastic model for concrete failure}},
  \bibinfo{journal}{International Journal of Solids and Structures}
  \bibinfo{volume}{43} (\bibinfo{year}{2006}) \bibinfo{pages}{7166--7196}.
\bibitem[{Grassl and Jir\'{a}sek(2006a)}]{GraJir06a}
\bibinfo{author}{P.~Grassl}, \bibinfo{author}{M.~Jir\'{a}sek},
  \bibinfo{title}{A plastic model with nonlocal damage applied to concrete},
  \bibinfo{journal}{International Journal for Numerical and Analytical Methods
  in Geomechanics} \bibinfo{volume}{30} (\bibinfo{year}{2006a})
  \bibinfo{pages}{71--90}.
\bibitem[{Grassl et~al.(2002)Grassl, Lundgren and Gylltoft}]{GraLunGyl02}
\bibinfo{author}{P.~Grassl}, \bibinfo{author}{K.~Lundgren},
  \bibinfo{author}{K.~Gylltoft}, \bibinfo{title}{Concrete in compression: a
  plasticity theory with a novel hardening law},
  \bibinfo{journal}{International Journal of Solids and Structures}
  \bibinfo{volume}{39} (\bibinfo{year}{2002}) \bibinfo{pages}{5205--5223}.
\bibitem[{Grassl and Rempling(2007)}]{GraRem07}
\bibinfo{author}{P.~Grassl}, \bibinfo{author}{R.~Rempling},
  \bibinfo{title}{Influence of volumetric-deviatoric coupling on crack
  prediction in concrete fracture tests}, \bibinfo{journal}{Engineering
  Fracture Mechanics} \bibinfo{volume}{74} (\bibinfo{year}{2007})
  \bibinfo{pages}{1683--1693}.
\bibitem[{Imran and Pantazopoulou(1996)}]{ImrPan96}
\bibinfo{author}{I.~Imran}, \bibinfo{author}{S.J. Pantazopoulou},
  \bibinfo{title}{Experimental study of plain concrete under triaxial stress},
  \bibinfo{journal}{\JACIM} \bibinfo{volume}{93} (\bibinfo{year}{1996})
  \bibinfo{pages}{589--601}.
\bibitem[{Jason et~al.(2006)Jason, Huerta, Pijaudier-Cabot and
  Ghavamian}]{JasHuePijGha06}
\bibinfo{author}{L.~Jason}, \bibinfo{author}{A.~Huerta},
  \bibinfo{author}{G.~Pijaudier-Cabot}, \bibinfo{author}{S.~Ghavamian},
  \bibinfo{title}{{An elastic plastic damage formulation for concrete:
  Application to elementary tests and comparison with an isotropic damage
  model}}, \bibinfo{journal}{Computer Methods in Applied Mechanics and
  Engineering} \bibinfo{volume}{195} (\bibinfo{year}{2006})
  \bibinfo{pages}{7077--7092}.
\bibitem[{Jir\'{a}sek and Ba\v{z}ant(2002)}]{JirBaz01}
\bibinfo{author}{M.~Jir\'{a}sek}, \bibinfo{author}{Z.P. Ba\v{z}ant},
  \bibinfo{title}{Inelastic Analysis of Structures}, \bibinfo{publisher}{John
  Wiley and Sons}, \bibinfo{address}{Chichester}, \bibinfo{year}{2002}.
\bibitem[{Jir\'{a}sek and Grassl(2008)}]{JirGra08}
\bibinfo{author}{M.~Jir\'{a}sek}, \bibinfo{author}{P.~Grassl},
  \bibinfo{title}{Evaluation of directional mesh bias in concrete fracture
  simulations using continuum damage models}, \bibinfo{journal}{Engineering
  Fracture Mechanics} \bibinfo{volume}{75} (\bibinfo{year}{2008})
  \bibinfo{pages}{1921--1943}.
\bibitem[{Jir\'{a}sek and Zimmermann(1998)}]{JirZim98}
\bibinfo{author}{M.~Jir\'{a}sek}, \bibinfo{author}{T.~Zimmermann},
  \bibinfo{title}{Rotating crack model with transition to scalar damage},
  \bibinfo{journal}{Journal of Engineering Mechanics} \bibinfo{volume}{124}
  (\bibinfo{year}{1998}) \bibinfo{pages}{277--284}.
\bibitem[{Ju(1989)}]{Ju89}
\bibinfo{author}{J.W. Ju}, \bibinfo{title}{On energy-based coupled
  elastoplastic damage theories: {C}onstitutive modeling and computational
  aspects}, \bibinfo{journal}{\IJSS} \bibinfo{volume}{25}
  (\bibinfo{year}{1989}) \bibinfo{pages}{803--833}.
\bibitem[{Kachanov(1980)}]{Kac80}
\bibinfo{author}{M.~Kachanov}, \bibinfo{title}{Continuum model of medium with
  cracks}, \bibinfo{journal}{Journal of the Engineering Mechanics Division}
  \bibinfo{volume}{106} (\bibinfo{year}{1980}) \bibinfo{pages}{1039--1051}.
\bibitem[{Karsan and Jirsa(1969)}]{KarJir69}
\bibinfo{author}{I.D. Karsan}, \bibinfo{author}{J.O. Jirsa},
  \bibinfo{title}{Behavior of concrete under compressive loadings},
  \bibinfo{journal}{\JSD} \bibinfo{volume}{95} (\bibinfo{year}{1969})
  \bibinfo{pages}{2543--2563}.
\bibitem[{Kormeling and Reinhardt(1982)}]{KorRei83}
\bibinfo{author}{H.A. Kormeling}, \bibinfo{author}{H.W. Reinhardt},
  \bibinfo{title}{Determination of the fracture energy of normal concrete and
  epoxy-modified concrete}, \bibinfo{type}{Stevin Laboratory}
  \bibinfo{number}{5-83-18}, Delft University of Technology,
  \bibinfo{year}{1982}.
\bibitem[{Kupfer et~al.(1969)Kupfer, Hilsdorf and {R\"{u}sch}}]{Kupfer69}
\bibinfo{author}{H.~Kupfer}, \bibinfo{author}{H.K. Hilsdorf},
  \bibinfo{author}{H.~{R\"{u}sch}}, \bibinfo{title}{{Behavior of concrete under
  biaxial stresses}}, \bibinfo{journal}{\JACI} \bibinfo{volume}{66}
  (\bibinfo{year}{1969}) \bibinfo{pages}{656--666}.
\bibitem[{Lee and Fenves(1998)}]{LeeFen98}
\bibinfo{author}{J.~Lee}, \bibinfo{author}{G.L. Fenves},
  \bibinfo{title}{Plastic-damage model for cyclic loading of concrete
  structures}, \bibinfo{journal}{\JEM} \bibinfo{volume}{124}
  (\bibinfo{year}{1998}) \bibinfo{pages}{892--900}.
\bibitem[{Leon(1935)}]{Leo35}
\bibinfo{author}{A.~Leon}, \bibinfo{title}{\"{U}ber die {S}cherfestigkeit des
  {B}etons}, \bibinfo{journal}{Beton und Eisen} \bibinfo{volume}{34}
  (\bibinfo{year}{1935}).
\bibitem[{Mazars(1984)}]{Mazars84}
\bibinfo{author}{J.~Mazars}, \bibinfo{title}{Application de la m\'{e}canique de
  l'endommagement au comportement non lin\'{e}aire et \`{a} la rupture du
  b\'{e}ton de structure}, \bibinfo{type}{{Th\`{e}se de Doctorat d'Etat}},
  Universit\'{e} Paris VI., \bibinfo{address}{France}, \bibinfo{year}{1984}.
\bibitem[{Mazars and Pijaudier-Cabot(1989)}]{MazPij89}
\bibinfo{author}{J.~Mazars}, \bibinfo{author}{G.~Pijaudier-Cabot},
  \bibinfo{title}{Continuum damage theory—-application to concrete},
  \bibinfo{journal}{Journal of Engineering Mechanics} \bibinfo{volume}{115}
  (\bibinfo{year}{1989}) \bibinfo{pages}{345}.
\bibitem[{Men\'{e}trey and Willam(1995)}]{Menetrey95}
\bibinfo{author}{P.~Men\'{e}trey}, \bibinfo{author}{K.J. Willam},
  \bibinfo{title}{{A triaxial failure criterion for concrete and its
  generalization}}, \bibinfo{journal}{\JACIS} \bibinfo{volume}{92}
  (\bibinfo{year}{1995}) \bibinfo{pages}{311--318}.
\bibitem[{Nguyen and Houlsby(2008)}]{NguHou08}
\bibinfo{author}{G.D. Nguyen}, \bibinfo{author}{G.T. Houlsby},
  \bibinfo{title}{A coupled damage--plasticity model for concrete based on
  thermodynamic principles: {Part I}: model formulation and parameter
  identification}, \bibinfo{journal}{International Journal for Numerical and
  Analytical Methods in Geomechanics} \bibinfo{volume}{32}
  (\bibinfo{year}{2008}) \bibinfo{pages}{353--389}.
\bibitem[{Nguyen and Korsunsky(2008)}]{NguKor08}
\bibinfo{author}{G.D. Nguyen}, \bibinfo{author}{A.M. Korsunsky},
  \bibinfo{title}{Development of an approach to constitutive modelling of
  {concrete: I}sotropic damage coupled with plasticity},
  \bibinfo{journal}{International Journal of Solids and Structures}
  \bibinfo{volume}{45} (\bibinfo{year}{2008}) \bibinfo{pages}{5483--5501}.
\bibitem[{Ortiz(1985)}]{Ort85}
\bibinfo{author}{M.~Ortiz}, \bibinfo{title}{A constitutive theory for the
  inelastic behavior of concrete}, \bibinfo{journal}{Mechanics of Materials}
  \bibinfo{volume}{4} (\bibinfo{year}{1985}) \bibinfo{pages}{67--93}.
\bibitem[{Papanikolaou and Kappos(2007)}]{PapKap07}
\bibinfo{author}{V.K. Papanikolaou}, \bibinfo{author}{A.J. Kappos},
  \bibinfo{title}{Confinement--sensitive plasticity constitutive model for
  concrete in triaxial compression}, \bibinfo{journal}{International Journal of
  Solids and Structures} \bibinfo{volume}{44} (\bibinfo{year}{2007})
  \bibinfo{pages}{7021--7048}.
\bibitem[{Patz\'{a}k(1999)}]{Pat99}
\bibinfo{author}{B.~Patz\'{a}k}, \bibinfo{title}{Object oriented finite element
  modeling}, \bibinfo{journal}{Acta Polytechnica} \bibinfo{volume}{39}
  (\bibinfo{year}{1999}) \bibinfo{pages}{99--113}.
\bibitem[{Patz\'{a}k and Bittnar(2001)}]{PatBit01}
\bibinfo{author}{B.~Patz\'{a}k}, \bibinfo{author}{Z.~Bittnar},
  \bibinfo{title}{Design of object oriented finite element code},
  \bibinfo{journal}{Advances in Engineering Software} \bibinfo{volume}{32}
  (\bibinfo{year}{2001}) \bibinfo{pages}{759--767}.
\bibitem[{Pivonka(2001)}]{Piv01}
\bibinfo{author}{P.~Pivonka}, \bibinfo{title}{Nonlocal plasticity models for
  localized failure}, \bibinfo{type}{Ph.{D.}\ thesis}, Technische Universit\"at
  Wien, Austria, \bibinfo{year}{2001}.
\bibitem[{Pramono and Willam(1989)}]{PraWil89}
\bibinfo{author}{E.~Pramono}, \bibinfo{author}{K.~Willam},
  \bibinfo{title}{Fracture energy-based plasticity formulation of plain
  concrete}, \bibinfo{journal}{\JEM} \bibinfo{volume}{115}
  (\bibinfo{year}{1989}) \bibinfo{pages}{1183--1203}.
\bibitem[{Resende(1987)}]{Res87}
\bibinfo{author}{L.~Resende}, \bibinfo{title}{A damage mechanics constitutive
  theory for the inelastic behaviour of concrete}, \bibinfo{journal}{\CMAME}
  \bibinfo{volume}{60} (\bibinfo{year}{1987}) \bibinfo{pages}{57--93}.
\bibitem[{S{\'a}nchez et~al.(2011)S{\'a}nchez, Huespe, Oliver, Diaz and
  Sonzogni}]{SanHueOli11}
\bibinfo{author}{P.~S{\'a}nchez}, \bibinfo{author}{A.~Huespe},
  \bibinfo{author}{J.~Oliver}, \bibinfo{author}{G.~Diaz},
  \bibinfo{author}{V.~Sonzogni}, \bibinfo{title}{A macroscopic damage-plastic
  constitutive law for modeling quasi-brittle fracture and ductile behavior of
  concrete}, \bibinfo{journal}{International Journal for Numerical and
  Analytical Methods in Geomechanics} \bibinfo{volume}{36}
  (\bibinfo{year}{2011}) \bibinfo{pages}{546--573}.
\bibitem[{Tao and Phillips(2005)}]{TaoPhi05}
\bibinfo{author}{X.~Tao}, \bibinfo{author}{D.V. Phillips}, \bibinfo{title}{A
  simplified isotropic damage model for concrete under bi-axial stress states},
  \bibinfo{journal}{Cement and Concrete Composites} \bibinfo{volume}{27}
  (\bibinfo{year}{2005}) \bibinfo{pages}{716--726}.
\bibitem[{Valentini and Hofstetter(2012)}]{ValHof12}
\bibinfo{author}{B.~Valentini}, \bibinfo{author}{B.V. Hofstetter},
  \bibinfo{title}{Review and enhancement of {3D} concrete models for
  large-scale numerical simulations of concrete structures},
  \bibinfo{journal}{International Journal for Numerical and Analytical Methods
  in Geomechanics}  (\bibinfo{year}{2012}). \bibinfo{note}{In press}.
\bibitem[{\v{C}ervenka and Papanikolaou(2008)}]{CerPap08}
\bibinfo{author}{J.~\v{C}ervenka}, \bibinfo{author}{V.K. Papanikolaou},
  \bibinfo{title}{Three dimensional combined fracture-plastic material model
  for concrete}, \bibinfo{journal}{International Journal of Plasticity}
  \bibinfo{volume}{24} (\bibinfo{year}{2008}) \bibinfo{pages}{2192--2220}.
\bibitem[{Voyiadjis and Kattan(2009)}]{VoyKat09}
\bibinfo{author}{G.Z. Voyiadjis}, \bibinfo{author}{P.I. Kattan},
  \bibinfo{title}{A comparative study of damage variables in continuum damage
  mechanics}, \bibinfo{journal}{International Journal of Damage Mechanics}
  \bibinfo{volume}{18} (\bibinfo{year}{2009}) \bibinfo{pages}{315--340}.
\bibitem[{Voyiadjis et~al.(2008)Voyiadjis, Taqieddin and Kattan}]{VoyTaqKat08}
\bibinfo{author}{G.Z. Voyiadjis}, \bibinfo{author}{Z.N. Taqieddin},
  \bibinfo{author}{P.I. Kattan}, \bibinfo{title}{Anisotropic damage--plasticity
  model for concrete}, \bibinfo{journal}{International Journal of Plasticity}
  \bibinfo{volume}{24} (\bibinfo{year}{2008}) \bibinfo{pages}{1946--1965}.
\bibitem[{Willam and Warnke(1974)}]{Willam74}
\bibinfo{author}{K.J. Willam}, \bibinfo{author}{E.P. Warnke},
  \bibinfo{title}{Constitutive model for the triaxial behavior of concrete},
  in: \bibinfo{booktitle}{Concrete Structures Subjected to Triaxial Stresses},
  volume~\bibinfo{volume}{19} of \textit{\bibinfo{series}{IABSE Report}},
  \bibinfo{publisher}{International Association of Bridge and Structural
  Engineers}, \bibinfo{address}{Zurich}, \bibinfo{year}{1974}, pp.
  \bibinfo{pages}{1--30}.

\end{thebibliography}
